\documentclass[10pt,twocolumn,twoside] {IEEEtran}
\ifCLASSINFOpdf
 \usepackage[pdftex]{graphicx}
\else
 \usepackage[dvips]{graphicx}
\fi
\usepackage{cite}
\usepackage{bm}
\usepackage{bbm}
\usepackage{amsmath,amssymb,amsfonts,lipsum}
\usepackage{float}
\usepackage{algorithmic}
\usepackage{graphicx}
\usepackage{textcomp}
\usepackage{color}
\usepackage{xcolor}
\usepackage{ragged2e}
\definecolor{light-gray}{gray}{0.92}
\usepackage{pifont}
\usepackage{colortbl}
\usepackage{makecell}
\usepackage{booktabs}
\usepackage{multirow}
\usepackage{algorithm}
\usepackage{stfloats}
\usepackage[caption=false, font=normalsize, labelfont=sf, textfont=sf]{subfig}
\def\BibTeX{{\rm B\kern-.05em{\sc i\kern-.025em b}\kern-.08em
		T\kern-.1667em\lower.7ex\hbox{E}\kern-.125emX}}

\usepackage[left=0.624in, right=0.624in, top=0.74in, bottom=1.04in]{geometry}
\usepackage{xpatch}
\makeatletter
\ExplSyntaxOn
\cs_new:Npn \bibColoredItems #1#2
  {
    \clist_map_inline:nn {#2} { \cs_new:cpn {bib@colored@##1} {#1} } 
  }
\ExplSyntaxOff
\newcommand\bib@setcolor[1]{%
  \ifcsname bib@colored@#1\endcsname
    \expanded{\noexpand\color{\csname bib@colored@#1\endcsname}}%
  \else
    \normalcolor
  \fi
}
\IfPackageLoadedTF{hyperref}{\@tempswatrue}{\@tempswafalse}
\if@tempswa
  \xpatchcmd\@bibitem {\H@item}{\bib@setcolor{#1}\H@item}{}{\PatchFailed}
  \xpatchcmd\@lbibitem{\H@item}{\bib@setcolor{#2}\H@item}{}{\PatchFailed}
\else
  \xpatchcmd\@bibitem {\item}  {\bib@setcolor{#1}\item}  {}{\PatchFailed}
  \xpatchcmd\@lbibitem{\item}  {\bib@setcolor{#2}\item}  {}{\PatchFailed}
\fi
\makeatother

\begin{document}
\title{Leveraging Deep Reinforcement Learning for Clustered Cell-Free Networking Over User Mobility}
\author{Ouyang~Zhou, \emph{Student Member,~IEEE}, Junyuan~Wang, \emph{Member,~IEEE}, Bo~Qian, \emph{Member,~IEEE}, \\Antonio~P\'{e}rez~Yuste, \emph{Senior Member,~IEEE} and Yusheng~Ji, \emph{Fellow,~IEEE}

\thanks{Received 14 September 2025; revised 18 March 2026; accepted 12 May 2026. This work was supported in part by National Natural Science Foundation of China under Grant 62371344, Fundamental Research Funds for the Central Universities, and JSPS Grant-in-Aid for Early-Career Scientists under Grant 25K21195. The work of A. P\'{e}rez Yuste was supported in part by a Seed Project Fund from UPM aimed at fostering cooperation with research groups in Asia, the United States, Canada, and Oceania. The work of Y. Ji was supported in part by JSPS KAKENHI Grant No. JP24K02937 and JST ASPIRE Grant No. JPMJAP2325. This paper was presented in part in IEEE Wireless Communications and Networking Conference (WCNC), Milan, Italy, March 2025 \cite{10978343}. The associate editor coordinating the review of this article and approving it for publication was M. Erol-Kantarci \textit{(Corresponding author: Junyuan Wang)}}
\thanks{O. Zhou is with the College of Electronic and Information Engineering, Tongji University, Shanghai 201804, China (e-mail: zoy@tongji.edu.cn).}
\thanks{J. Wang is with the College of Electronic and Information Engineering, Shanghai Institute of Intelligent Science and Technology, and the Institute of Advanced Study, Tongji University, Shanghai, 201804, China (e-mail: junyuanwang@tongji.edu.cn).}
\thanks{B. Qian is with the Graduate School of Information Science and Technology, The University of Tokyo, Tokyo 113-8657, Japan (e-mail: boqian@ieee.org).}
\thanks{A. P\'{e}rez Yuste is with the Department of Communications and Audio and Video Engineering, Technical University of Madrid, Campus Sur UPM, 28031 Madrid, Spain (e-mail: antonio.perez@upm.es).}
\thanks{Y. Ji is with National Institute of Informatics, Tokyo 101-8430, Japan (e-mail: kei@nii.ac.jp).}
}
\maketitle

\begin{abstract}
Clustered cell-free networking paves a new way for enabling scalable joint transmission among access points (APs) by partitioning the whole network into non-overlapping subnetworks. Previous works adopted clustering algorithms, graph partitioning methods or conventional continuous optimization theories to partition a network based on the channels between all users and all APs, resulting in huge channel measurement and computational costs. This makes these methods difficult to be implemented in practical systems since the optimal network partition could vary frequently due to user mobility. In addition, existing methods were usually designed for specific clustered cell-free networking problems with different optimization algorithms employed. In this paper, we leverage deep reinforcement learning (DRL) for clustered cell-free networking so as to rapidly adapt to user movements in dynamic environments, and propose a deep deterministic policy gradient based clustered cell-free networking (DDPG-C$^{2}$F) framework that can be adapted in various application scenarios. 
Moreover, in our framework, only one single channel needs to be estimated at each AP as the input of the neural network, which greatly reduces the channel measurement costs for clustered cell-free networking, and the training and inference costs of our framework. The proposed DDPG-C$^{2}$F framework is then applied to various clustered cell-free networking problems with different objectives and constraints to demonstrate its performance. Simulation results show that our framework outperforms existing baselines in all scenarios. Moreover, we show that the proposed framework can reduce the handover cost over user mobility, and is robust to dynamic scenarios with random user joining or leaving.
\end{abstract}


\section{Introduction}
To meet the extremely high data rate and low delay requirements of numerous intelligent services, such as autonomous driving, digital twins and holographic communications, base-stations (BSs) would be ultra-densely deployed in the sixth-generation (6G) mobile communication systems \cite{10158439}. However, ultra-dense deployment of BSs leads to escalated cell-edge problem in current cellular networks, i.e., a large number of users would be located in the cell boundary areas and experience poor services due to strong inter-cell interference. 

To avoid the severe cell-edge problem, one solution is to coordinate all geographically distributed access points (APs) (which can be regarded as mini BSs) to jointly serve all users on the same time-frequency resources. Such a full cooperation idea has been adopted in distributed antenna systems (DASs) \cite{7031453}, network multiple-input-multiple-output (MIMO) \cite{6095627}, cloud radio access network (C-RAN) \cite{pan} and cell-free massive MIMO \cite{9650567}. It has been demonstrated that coordinating all APs can achieve seamless coverage and high capacity \cite{9650567}, which, however, leads to unaffordable joint processing complexity and signaling overhead. To address these issues, it was proposed to serve each user only by its nearby APs that dominate the channel \cite{6783666, 8244310, 7317799}. Yet the serving AP sets of different users might overlap and their signal processing could be coupled with each other, complicating data transmission. 

In order to alleviate the cell-edge problem and provide universal services to users by facilitating joint processing of APs with tolerable complexity, a novel clustered cell-free network architecture was proposed in \cite{8007415, 2022, JunyuanTWC}. Specifically, the whole network is dynamically partitioned into a number of non-overlapping parallel operating subnetworks with joint processing of APs performed in each subnetwork independently. To efficiently eliminate the mutual interference via joint processing, users and APs strongly interfering with each other should be assigned into the same subnetwork. How to group users and APs to form subnetworks is a key problem, which is referred to as clustered cell-free networking \cite{JunyuanTWC}. 

AP-centric clustering strategies were proposed in \cite{8849663} and \cite{10535986}, where APs are first clustered, followed by the association of users with these AP clusters. In \cite{8849663}, a hierarchical clustering algorithm was used to group APs, whereas a Gaussian mixture model (GMM) was employed in \cite{10535986}. User-centric clustering schemes were explored in \cite{8644255} and \cite{10458891}, where users are first grouped using K-means \cite{8644255} or hierarchical clustering \cite{10458891}, and then APs are associated with these user groups. Instead of treating the clustered cell-free networking problem as a two-stage clustering problem, \cite{8007415, 2022, JunyuanTWC, 10622456, 10776991} formulated it as a graph partitioning problem and employed spectral clustering to solve it. However, the aforementioned clustered cell-free networking schemes often lead to imbalanced subnetworks when users are congregated at some hot spots. In this case, the signaling overhead and joint processing complexity in the largest subnetwork could be still unaffordable. To constrain subnetwork sizes, \cite{10206557, 10619648, 10766356} proposed to bound the number of users in each subnetwork, while \cite{10973152} balanced the number of APs across all subnetworks. With these constraints, the clustered cell-free networking problem was studied by relaxing it into a continuous solvable form, which usually leads to suboptimal solutions.

Despite these achievements, most of the above works considered a static case with fixed users. However, users move frequently in practical systems, which brings additional challenges to clustered cell-free networking. In fact, as users move, the optimal network partition might vary. To maintain optimal clustered cell-free networking by adapting to the varying network topology, channels between users and APs need to be measured frequently and the clustered cell-free networking algorithm needs to be performed repeatedly. As the channel state informations (CSIs) between all users and all APs are required by the existing clustering-based algorithms \cite{8849663, 10535986, 8644255, 10458891}, the graph partitioning-based algorithms \cite{8007415, 2022, JunyuanTWC, 10622456, 10776991} and the classical optimization-based algorithms \cite{10206557, 10619648, 10766356, 10973152}, adopting these methods would lead to huge channel measurement costs. Moreover, these approaches are usually of high complexity, and thus might not be able to produce the optimal network partition promptly in high-mobility scenarios.

Deep reinforcement learning (DRL) has been widely explored in real-time management problems, such as resource allocation \cite{10278100, 10498103, 9745785, 9751765, 10570789} and user association \cite{9736948, 9625449, 9849036, 10496819, 10018462, 10225848, 10144362}. In reinforcement learning (RL), an agent learns to make sequential decisions to adapt to the varying environment through trial and error, while DRL enhances RL by utilizing deep neural networks (DNNs) to automatically adjust its strategy with low computational complexity. DRL can be applied to solve both continuous and discrete control tasks. Power control is one of the most classical continuous control problems. DRL-based approaches were proposed for uplink transmit power control in \cite{10278100, 10498103} and downlink power allocation in \cite{9745785} to maximize the overall rate performance, where deep deterministic policy gradient (DDPG), a popular DRL algorithm for continuous control, was employed. It was demonstrated that DRL-based approaches significantly reduce the computational complexity while approaching or even outperforming conventional optimization methods. Apart from power control, DRL has also been explored for other continuous control tasks, such as distributed beamforming \cite{9751765, 10570789}.

Compared to continuous control tasks, utilizing DRL for discrete control with high-dimensional solution space is more challenging. The DRL algorithms for discrete control, such as deep Q-network (DQN), produce the optimal solution through an exhaustive search over the solution space, thereby suffering from the curse of dimensionality. For instance, \cite{9736948} developed a DQN-based clustering method to group users and APs into subnetworks by evaluating all possible partitions, which is, however, infeasible for a large-scale network since the number of possible partitions grows exponentially as the network scales. To reduce the computational complexity, \cite{9625449, 9849036, 10496819} transformed discrete user association problems into continuous ones, and employed DDPG method to select serving APs for each user by optimizing the relaxed continuous association indicators and then discretizing them. \cite{10018462, 10225848, 10144362} applied multi-agent reinforcement learning (MARL) algorithm for user association, where each AP acts as an agent and determines the user association independently. Although this distributed method significantly reduces the computational complexity due to the parallel operation of multiple agents with a low-dimensional solution space, running DRL algorithms on massive APs could incur huge energy consumption.

Despite the above successes in leveraging DRL for user association, these approaches assumed that the serving AP sets of different users can overlap. By sharp contrast, clustered cell-free networking requires users and APs to be grouped into nonoverlapping subnetworks, posing a significant challenge for developing DRL-based approaches. This is due to the fact that DRL usually utilizes the posterior penalty to regularize the agent to meet the requirements rather than posing hard constraints during the learning process. How to explore DRL for clustered cell-free networking still remains largely unknown.

In this paper, we leverage DRL to tackle the clustered cell-free networking problem for a large-scale wireless network over user mobility. The main contributions of this paper are summarized as follows:
\begin{itemize}
\item We propose a new performance metric, the balance of subnetworks, as a function of the number of users and the number of APs in each subnetwork. Simulation results corroborate that the subnetwork sizes can be effectively balanced by maximizing the defined balance metric.
\item We formulate a general clustered cell-free networking problem over a time horizon of multiple time intervals, which can be tailored to accommodate different practical requirements in different scenarios. To solve this general problem efficiently, a novel deep deterministic policy gradient based clustered cell-free networking (DDPG-C$^{2}$F) framework is proposed. Particularly, instead of directly learning the network partition, our proposed framework learns the features of a number of subnetwork anchors to avoid the curse of dimensionality and then affiliate users and APs with the anchors to form non-overlapping subnetworks.
\item In contrast to existing methods that require the CSIs between all users and all APs, the proposed  DDPG-C$^{2}$F framework only requires the large-scale fading coefficient of the strongest channel between each AP and users as input, which significantly reduces the channel measurement costs yet without any performance degradation, as verified by simulations. In addition, this design reduces the dimensionality of the state space, and thus accelerates both training and inference of our DDPG-C$^{2}$F framework.
\item We apply the proposed DDPG-C$^{2}$F framework to four representative clustered cell-free networking problems with different objectives and constraints to demonstrate its effectiveness and generality. Simulation results show that the proposed DDPG-C$^{2}$F framework achieves significant performance gains over the state-of-the-art benchmarks in all scenarios.
\item Further simulations show that the proposed DDPG-C$^{2}$F framework incurs low handover cost over user mobility, since it learns similar anchor features over consecutive time intervals. Moreover, as both the state space and the action space
are independent of the number of users, our framework does not need to learn from scratch when users randomly access or leave the network.
\end{itemize}

The rest of this paper is organized as follows. Section II introduces the system model and formulates the problem. The DRL-based clustered cell-free networking framework is proposed in Section III, followed by case studies with rate- and energy efficiency-oriented problems in Section IV and Section V, respectively. Discussions are provided in Section VI, and concluding remarks are summarized in Section VII.

\begin{figure}[t]
\centering               
\includegraphics[width=0.4\textwidth]{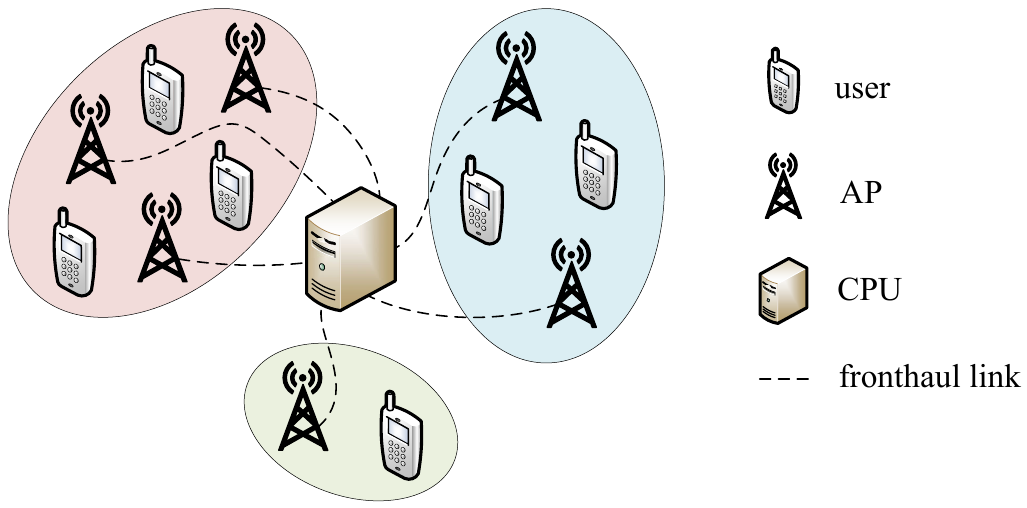}
\caption{Graphical illustration of clustered cell-free networking.}
\end{figure}

\begin{figure*}[!hb]
\begin{normalsize} 
\hrulefill
\begin{equation}
R_{k}=\log_{2}\left(1+\frac{|\mathbf{g}_{k, \mathcal{B}_{m}} \mathbf{w}_{k, \mathcal{B}_{m}}|^{2} P_{k}}{\sum_{u_{j} \in \mathcal{U}_{m}, j \neq k} |\mathbf{g}_{k, \mathcal{B}_{m}} \mathbf{w}_{j, \mathcal{B}_{m}}|^{2}P_{j}+\sum_{n =1, n \neq m}^{M}\sum_{u_{j} \in \mathcal{U}_{n}} |\mathbf{g}_{k, \mathcal{B}_{n}} \mathbf{w}_{j, \mathcal{B}_{n}}|^{2}P_{j}+N_{0}}\right)\tag{6}.
\end{equation}
\end{normalsize}
\end{figure*}

Throughout this paper, $*$ denotes the conjugate operator. $\mathbb{E}[\cdot]$ is the expectation operator. $\triangleq$ represents the definition operator. $\|\cdot\|$ stands for the L2 norm of a vector. $x \sim \mathcal{C} \mathcal{N}\left(u, \sigma^{2}\right)$ and $x \sim \mathcal{N}\left(u, \sigma^{2}\right)$ denote a complex and a real Gaussian random variable, respectively, with mean $u$ and variance $\sigma^{2}$. $|\mathcal{X}|$ represents the cardinality of set $\mathcal{X}$.

\section{System Model and Problem Formulation}
Consider a downlink large-scale wireless network with $L$ geographically distributed single-antenna APs and $K$ single-antenna users. The APs are connected to a central processing unit
(CPU) via fronthaul links. The set of users and the set of APs are denoted by $\mathcal{U}=\left\{u_{1}, u_{2}, \cdots, u_{K}\right\}$ and $\mathcal{B}=\left\{b_{1}, b_{2}, \cdots, b_{L}\right\}$, respectively, with $\left|\mathcal{U}\right|={K}$ and $\left|\mathcal{B}\right|={L}$. With clustered cell-free networking, as illustrated in Fig.~1, the network is partitioned into $M$ nonoverlapping subnetworks, in each of which the APs jointly serve the users therein under time-division duplex (TDD) operation. Let $\mathcal{U}_{m}$ and $\mathcal{B}_{m}$ denote the set of users and the set of APs in the $m$th subnetwork with $\left|\mathcal{U}_{m}\right|={K}_{m}$ and $\left|\mathcal{B}_{m}\right|={L}_{m}$, respectively. The network partition is denoted by $\mathcal{C}=\left\{\mathcal{C}_{1}, \mathcal{C}_{2}, \cdots, \mathcal{C}_{M}\right\}$ with
\begin{equation}
\mathcal{C}_{m}=\mathcal{U}_{m}\cup\hspace{1mm}\mathcal{B}_{m}, \forall m,
\end{equation}
and
\begin{equation}
\mathcal{C}_{m}\cap\mathcal{C}_{n}=\emptyset, \forall n\neq m.
\end{equation}

\subsection{Signal Model}
Consider user $u_{k}\in\mathcal{U}_{m}$ in the $m$th subnetwork as the reference user. The received signal $y_{k}$ at
user $u_{k}$ is given by
\begin{align}
y_{k}=&\underbrace{\mathbf{g}_{k, \mathcal{B}_{m}} \mathbf{x}_{k, \mathcal{B}_{m}}}_{\text {desired signal }}
\hspace{1mm}+\underbrace{\sum_{u_{j} \in \mathcal{U}_{m}, j \neq k} \mathbf{g}_{k, \mathcal{B}_{m}} \mathbf{x}_{j, \mathcal{B}_{m}}}_{\text {intra-subnetwork interference }}\notag \\
&+\underbrace{\sum_{n =1, n \neq m}^{M}\sum_{u_{j} \in \mathcal{U}_{n}} \mathbf{g}_{k, \mathcal{B}_{n}} \mathbf{x}_{j, \mathcal{B}_{n}}}_{\text {inter-subnetwork interference }}+\hspace{1mm}z_{k},
\end{align}
where $\mathbf{x}_{k, \mathcal{B}_{m}}\in \mathbb{C}^{L_{m} \times 1}$
denotes the transmitted signal from the APs in $\mathcal{B}_{m}$ to user $u_{k}$.
$z_{k} \sim \mathcal{C N}\left(0, N_{0}\right)$ represents the additive white Gaussian noise (AWGN) with zero mean and variance $N_{0}$.
$\mathbf{g}_{k, \mathcal{B}_{m}} \in \mathbb{C}^{1 \times L_{m}}$
is the channel gain vector from the APs in $\mathcal{B}_{m}$ to user $u_{k}$, given by
\begin{equation}
\mathbf{g}_{k, \mathcal{B}_{m}}=\boldsymbol{\gamma}_{k, \mathcal{B}_{m}} \circ \hspace{1mm}\mathbf{h}_{k, \mathcal{B}_{m}},
\end{equation}
where $\circ$ is the Hadamard product. $\boldsymbol{\gamma}_{k, \mathcal{B}_{m}} \in \mathbb{R}^{1 \times L_{m}}$
denotes the corresponding large-scale fading vector. $\mathbf{h}_{k, \mathcal{B}_{m}} \in \mathbb{C}^{1 \times L_{m}}$
represents the small-scale fading vector with entries modeled as independent and identically distributed (i.i.d) 
complex Gaussian random variables with zero mean and unit variance.

By assuming that some linear spatial precoding scheme is adopted in each subnetwork for intra-subnetwork interference mitigation, the transmitted signal from the APs in $\mathcal{B}_{m}$ to user $u_{k}\in\mathcal{U}_{m}$, 
$\mathbf{x}_{k, \mathcal{B}_{m}}$, is written as
\begin{equation}
\mathbf{x}_{k, \mathcal{B}_{m}}=\mathbf{w}_{k, \mathcal{B}_{m}} \cdot\hspace{1mm} s_{k},
\end{equation}
where $\mathbf{w}_{k, \mathcal{B}_{m}}\in \mathbb{C}^{L_{m} \times 1}$ denotes the precoding vector with $\left\|\mathbf{w}_{k, \mathcal{B}_{m}}\right\|=1$. $s_{k}$ represents the information-bearing signal for user $u_{k}$ with $\mathbb{E}[s_{k}s_{k}^{*}]=P_{k}$, where $P_{k}$ is the transmit power allocated to user $u_{k}$. 
By normalizing the total system bandwidth into unity,
the achievable rate $R_{k}$ of user $u_{k} \in \mathcal{U}_{m}$ can be obtained as (6), shown at the bottom of the previous page.

\subsection{Balance of Subnetworks}
To mitigate the intra-subnetwork interference, the CSIs between users and APs in each subnetwork are required to perform spatial precoding, leading to huge channel measurement overhead and high joint processing complexity in a large subnetwork, both of which are undesirable in practice. Therefore, balancing subnetwork sizes to avoid large subnetworks is of great importance for clustered cell-free networking.

As the channel measurement overhead and joint processing complexity increase with both the number of users and the number of APs in a subnetwork, in this paper, we define the balance of subnetworks as
\begin{equation}
\rho=\rho_{u}\cdot\hspace{1mm}\rho_{b}\tag{7},
\end{equation}
where $\rho_{u}$ and $\rho_{b}$ measure the balance of the number of users and the number of APs across all subnetworks, respectively. Inspired by the min-max ratio approach in \cite{jain1984quantitative}, $\rho_{u}$ and $\rho_{b}$ are defined as
\begin{equation}
\rho_{u}=\frac{\mathrm{min}\left\{K_{1}, K_{2}, \cdots, K_{M}\right\}}{\mathrm{max}\left\{K_{1}, K_{2}, \cdots, K_{M}\right\}}\tag{8},
\end{equation}
and
\begin{equation}
\rho_{b}=\frac{\mathrm{min}\left\{L_{1}, L_{2}, \cdots, L_{M}\right\}}{\mathrm{max}\left\{L_{1}, L_{2}, \cdots, L_{M}\right\}}\tag{9},
\end{equation}
respectively. It can be easily seen that the balance of subnetworks $\rho$ is maximized when both users and APs are evenly allocated to $M$ subnetworks. In this case, the maximum number of channels that need to be measured in a single subnetwork is minimized, as we will show later in Figs. 6(a)--(b) and Figs. 11(a)--(b) that maximizing the balance of subnetworks with our proposed DDPG-C$^{2}$F framework results in smaller maximum number of measured channels across all subnetworks compared to other benchmarks.

\subsection{Problem Formulation}
As the optimal network partition could vary as users move, we aim to optimize the long-term performance over user mobility in this paper rather than focusing on a static scenario. Specifically, we consider a time horizon with $T$ time intervals, and each user's position may change at the end of each time interval along with its moving speed and direction.
Let $\mathcal{C}^{(t)}$ denote the network partition at time interval $t$. Then a general long-term clustered cell-free networking problem over the time horizon of $T$ intervals can be formulated as
\begin{align}
\mathcal{P}1:\hspace{3mm}\mathop{\max}_{\mathcal{C}^{(t)}=\left\{\mathcal{C}_{1}^{(t)}, \mathcal{C}_{2}^{(t)}, \cdots, \mathcal{C}_{M}^{(t)}\right\}}&~\frac{1}{T}\sum_{t=1}^{T}f\left(\mathcal{C}^{(t)}\right)\tag{10}\\
\text{s.t.}&~\mathcal{C}_{m}^{(t)}\cap\mathcal{C}_{n}^{(t)}=\emptyset, \forall n\neq m\tag{11},\\
&~\mathcal{C}_{m}^{(t)}\cap\mathcal{U}\neq\emptyset,
\forall m\tag{12},\\
&~\mathcal{C}_{m}^{(t)}\cap\mathcal{B}\neq\emptyset,
\forall m\tag{13},\\
&~\bigcup\limits_{m=1}^{M}\mathcal{U}_{m}^{(t)}=\mathcal{U}\tag{14},\\
&~\mathcal{C}^{(t)} \in \mathcal{V}\tag{15},
\end{align}
where $f\left(\mathcal{C}^{(t)}\right)$ is the general objective function that can be customized to facilitate various optimization goals. Four case studies will be presented in Sections IV and V. (11) indicates that the network is decomposed into nonoverlapping subnetworks. (12) and (13) require that each subnetwork contains at least one user and one AP to avoid trivial subnetworks. (14) ensures that all users are allocated to $M$ subnetworks to avoid service outage. (15) is a general constraint representing the requirements imposed in the specific scenario with $\mathcal{V}$ collecting the requirements. 

Clearly, problem $\mathcal{P}1$ is a combinatorial optimization problem, which is hard to solve in polynomial time. Moreover, the objective function and constraint (15) are both general, while utilizing conventional optimization methods requires to specify the problem and apply different mathematical optimization techniques in different cases. In addition, to promptly update the network partition as users move, conventional optimization algorithms need to be repeatedly run, and all the channels between users and APs need to be frequently estimated, leading to prohibitively huge channel measurement and computational costs. In this paper, we will leverage DRL to tackle these issues and propose a general low-cost framework for clustered cell-free networking.

\section{Deep Reinforcement Learning Framework for Clustered Cell-Free Networking}
In DRL, a goal-oriented agent learns the optimal policy through trial and error, which requires no knowledge about the environmental dynamics. Thanks to this model-free nature, we will explore DRL to solve the general clustered cell-free networking problem $\mathcal{P}1$, and propose a novel deep deterministic policy gradient based clustered cell-free networking (DDPG-C$^{2}$F) framework. This framework can be easily tailored to accommodate different objective functions and constraints, and notably, only a small number of channel estimates are required as the input of the framework for clustered cell-free networking optimization.

\subsection{Motivation of DDPG}
Problem $\mathcal{P}1$ is a sequential decision-making problem, which can be modeled as a Markov decision process (MDP) \cite{8714026} characterized by state space $\mathcal{S}$, action space $\mathcal{A}$, and reward space $\mathcal{R}$. Specifically, the CPU can function as an agent, which observes the current state $s^{(t)}\in \mathcal{S}$ of the network environment and selects an action $a^{(t)}\in \mathcal{A}$ to partition the whole network into $M$ subnetworks. The agent then receives the corresponding reward $r^{(t)}\in \mathcal{R}$ and the environment transits to the next state $s^{(t+1)}$. We denote $\pi$ as the policy of the agent, which guides action selection during the interaction with the environment. The goal of the agent is to find the optimal policy $\pi^{*}$ that maximizes the expected cumulative reward, i.e.,
\begin{equation}
\pi^{*}=\arg\mathop{ \max}_{\pi}\mathbb{E}_{\pi}\left[\sum_{t=1}^{T}\zeta^{t-1}r^{(t)}\right]\tag{16},
\end{equation}
where $\zeta\in\left(0, 1\right)$ is the discount factor for future rewards. $\zeta\rightarrow 0$ indicates that the agent prioritizes the immediate reward, while $\zeta\rightarrow 1$ represents the preference on the long-term reward. 

DRL has been shown to be an effective tool to solve MDP problems in many areas. Several popular algorithms have been proposed, which can be classified as discrete or continuous according to whether the action space is discrete or continuous. Discrete optimization algorithms, such as DQN, find the optimal policy through an exhaustive search over the discrete action space. Intuitively, our clustered cell-free networking problem should be addressed by discrete optimization algorithms since it is combinatorial. However, there are $M^{K+L}$ possible network partitions in total, resulting in an extremely large action space even for a moderate number of users $K$ and a moderate number of APs $L$. This makes discrete optimization algorithms difficult to converge. Therefore, we transform our problem $\mathcal{P}1$ into a continuous decision problem, and resort to the continuous DDPG algorithm.\footnote{Note that other continuous DRL algorithms, such as twin delayed deep deterministic policy gradient (TD3) and proximal policy optimization (PPO), can also be employed in the proposed DRL-based framework to train the agent. In this paper, we simply adopt DDPG to demonstrate that the proposed framework can tackle the difficulties in leveraging DRL for clustered cell-free networking, and reduces the signaling and computational costs of implementing clustered cell-free networking in practice. Adopting an advanced DRL algorithm and/or further improving the adopted algorithm in the proposed framework could lead to better performance, which would be carefully studied in the future.}

\subsection{Architecture of Proposed DDPG-C$^{2}$F}
To facilitate DDPG for clustered cell-free networking, instead of directly learning the network partition, we propose to learn the features of $M$ subnetwork anchors,\footnote{Here, the anchor of a subnetwork can be seen as the unique representation of that subnetwork.} and then affiliate each user and AP with one of the anchors to form subnetworks. The state, action and reward of the proposed DDPG-C$^{2}$F framework are defined as follows.

\begin{figure*}[t]
\centering                      
\includegraphics[width=0.9\textwidth]{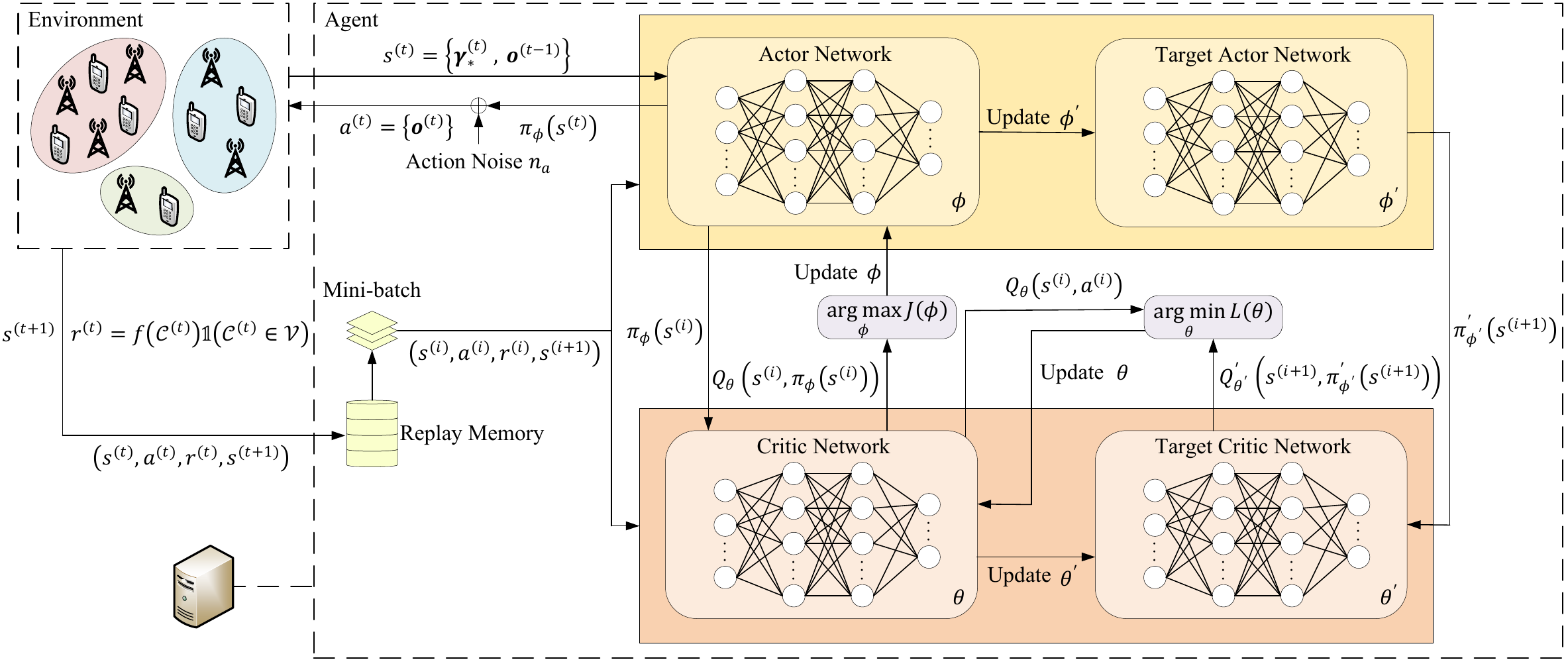}
\caption{Detailed architecture of the proposed DDPG-C$^{2}$F framework.}
\end{figure*}

\begin{itemize}
\item{\textbf{State:}} Note that the optimal network partition depends on the channel conditions between APs and users. In order to avoid the effect of fast-varying small-scale fading on the networking result, only large-scale fading is utilized for clustered cell-free networking. However, the large-scale fading coefficients between all APs and all users need to be estimated. To reduce the channel estimation cost, we propose to utilize the highest large-scale fading coefficient between an AP and all users only as the channel feature of this AP. We will show later in Sections IV-B and V-B that employing this simple channel feature in our DDPG-C$^{2}$F framework brings no performance degradation compared to inputting all the large-scale fading coefficients between APs and users. Specifically, the channel feature of AP $b_{l}$ at time interval $t$ is set as
\begin{equation}
\gamma_{l^{*}}^{(t)}=\mathrm{max}\left\{\gamma_{1, l}^{(t)}, \gamma_{2, l}^{(t)}, \cdots, \gamma_{K, l}^{(t)}\right\}\tag{17},
\end{equation}
where $\gamma_{k, l}^{(t)}$ is the large-scale fading coefficient from AP $b_{l}$ to user $u_{k}$.\footnote{In the TDD mode, downlink CSIs can be easily obtained by exploiting the uplink-downlink channel reciprocity.} Notably, with this setting, each AP only needs to estimate the strongest channel, which greatly reduces the computational cost of channel estimation. The state space can be then formalized as
\begin{equation}
s^{(t)}=\left\{\gamma_{*}^{(t)}, o^{(t-1)}\right\}\tag{18},
\end{equation}
where $\gamma_{*}^{(t)}\hspace{-1mm}=\hspace{-1mm}\left\{\hspace{-0.5mm}\gamma_{1^{*}}^{(t)}\hspace{-0.5mm}, \gamma_{2^{*}}^{(t)}\hspace{-0.5mm}, \cdots\hspace{-0.5mm}, \gamma_{L^{*}}^{(t)}\hspace{-0.5mm}\right\}$ collects $L$ highest large-scale fading coefficients corresponding to $L$ APs at time interval $t$, and $o^{(t-1)}=\left\{\boldsymbol{{o}}_{1}^{(t-1)}, \boldsymbol{{o}}_{2}^{(t-1)}, \cdots, \boldsymbol{{o}}_{M}^{(t-1)}\right\}$ denotes the set of the features of $M$ subnetwork anchors at time interval $t-1$. It can be seen that employing $\gamma_{*}^{(t)}$ instead of all the large-scale fading coefficients between APs and users reduces the dimensionality of the state space, which would accelerate the training and inference of the proposed framework. Moreover, the previous anchor features at time interval $t-1$ are incorporated into the state space to ensure that the proposed DDPG-C$^{2}$F framework strictly follows the MDP, and thus can converge with sufficient training \cite{singh2000convergence}.

\item{\textbf{Action:}} The action space corresponds to the feature set of $M$ subnetwork anchors at time interval $t$, which is defined as
\begin{equation}
a^{(t)}=\left\{o^{(t)}\right\}\tag{19}.
\end{equation}
Note that an anchor's feature could include its location as well as other essential parameters. We will show later in Sections IV-A and V-A that by customizing the anchor feature, the proposed DDPG-C$^{2}$F framework can be applied in different scenarios.

\item{\textbf{Reward:}} Following the objective function in (10) and the constraint in (15), the reward function is defined as
\begin{equation}
r^{(t)}=f\left(\mathcal{C}^{(t)}\right)\cdot\mathbbm{1}\hspace{-0.5mm}\left(\mathcal{C}^{(t)} \in \mathcal{V}\right)\tag{20},
\end{equation}
where $\mathbbm{1}\left(x\right)$ is the indicator
function. $\mathbbm{1}\left(x\right)=1$ if condition $x$ is true, and $\mathbbm{1}\left(x\right)=0$ otherwise. Here, the indicator function serves as a penalty term to regularize the output to meet the case-specific constraint (15). We only include constraint (15) in the penalty term because the proposed DDPG-C$^{2}$F framework has already fostered the agent to meet the constraints in (11)--(14). Specifically, constraints (11) and (14) are satisfied as each user/AP is allowed to affiliate with only one of the subnetwork anchors. In addition, we consider the balance of subnetworks $\rho^{(t)}$ as an important performance metric in this paper. By incorporating it into the objective function $f\left(\mathcal{C}^{(t)}\right)$, as shown later in Sections IV-A and V-A, the agent would learn to produce balanced subnetworks, i.e., the requirements in (12) and (13) would be met implicitly.
\end{itemize} 

\begin{algorithm}[t]
\caption{Training Process of DDPG-C$^{2}$F framework}
\begin{algorithmic}[1] 
\STATE {Initialize the actor network $\pi_{\phi}(s)$ and 
the critic network $Q_{\theta}(s, a)$ with weights $\phi$ and $\theta$}.
\STATE {Initialize the target actor network $\pi^{'}_{\phi^{'}}(s)$ and the target critic network $Q^{'}_{\theta^{'}}(s, a)$ with $\phi^{'}=\phi$ and $\theta^{'}=\theta$}.
\STATE {Initialize the replay buffer with capacity $D$.}
\FOR{$e=1$ to $E$}
\STATE {Set the initial state $s^{(1)}$ and update the action noise $n_{a}$}.
\FOR{$t=1$ to $T$}
\STATE {Select the action $a^{(t)}=\pi_{\phi}(s^{(t)})+n_{a}$}.
\STATE {Obtain the reward $r^{(t)}$ and the next state $s^{(t+1)}$.}
\STATE {Replace the old experience by new experience $\left(s^{(t)}, a^{(t)}, r^{(t)}, s^{(t+1)}\right)$}.
\STATE {Sample a mini-batch of interaction experiences $\left(s^{(i)}, a^{(i)}, r^{(i)}, s^{(i+1)}\right)$ with size $B$ randomly}.
\STATE {Calculate the target value: 

$y^{(i)}=r^{(i)}+\zeta Q^{'}_{\theta^{'}}(s^{(i+1)}, \pi^{'}_{\phi^{'}}(s^{(i+1)}))$}.
\STATE {Update critic weights $\theta$ with gradient descent method to minimize the loss function: 

$L(\theta)=\frac{1}{B}\sum_{i}(y^{(i)}-Q_{\theta}(s^{(i)}, a^{(i)}))^{2}$}.
\STATE {Update actor weights $\phi$ with gradient ascent method to maximize the action-value function:

$J(\phi)=\frac{1}{B}\sum_{i}Q_{\theta}(s^{(i)}, \pi_{\phi}(s^{(i)}))$}.
\STATE {Update the target network weights $\theta^{'}$ and $\phi^{'}$:

$\theta^{'}\leftarrow\delta\theta+(1-\delta)\theta^{'}$,

$\phi^{'}\leftarrow\delta\phi+(1-\delta)\phi^{'}$}.
\ENDFOR
\ENDFOR
\end{algorithmic} 
\end{algorithm}

\subsection{Training Process of Proposed DDPG-C$^{2}$F}
DDPG finds the optimal policy $\pi^{*}$ by employing separate actor and critic networks. Fig. 2 illustrates the training process of the proposed DDPG-C$^{2}$F framework. The actor network learns a policy function $\pi_{\phi}(s)$ to generate an action $a$ with weight $\phi$, and the critic network exploits a Q-function $Q_{\theta}(s, a)$ to evaluate the expected cumulative reward of action $a$ with weight $\theta$. As the agent aims to maximize the expected cumulative reward, the objective function for training the actor network $\pi_{\phi}(s)$ can be expressed as
\begin{equation}
J\left(\phi\right)=\mathbb{E}\left[Q_{\theta}\left(s^{(t)}, \pi_{\phi}(s^{(t)})\right)\right]\tag{21},
\end{equation}
where $\pi_{\phi}(s^{(t)})=a^{(t)}$. The actor network's parameter $\phi$ can be then updated by the gradient ascent method with
\begin{equation}
\phi\leftarrow\phi+\alpha_{\phi}\nabla_{\phi}J\left(\phi\right)\tag{22},
\end{equation}
where $\alpha_{\phi}$ denotes the learning rate of the actor network.

As for the critic network, it needs to estimate the true value of $Q_{\theta}\left(s^{(t)},\pi_{\phi}(s^{(t)})\right)$. The loss function for training the critic network $Q_{\theta}\left(s, a\right)$ can be expressed as 
\begin{equation}
L(\theta)=\mathbb{E}\left[(y^{(t)}-Q_{\theta}(s^{(t)}, a^{(t)}))^{2}\right]\tag{23},
\end{equation}
where $y^{(t)}$ denotes the temporal-difference (TD) target for $Q_{\theta}\left(s^{(t)},a^{(t)}\right)$. To ensure the stability of the training process, DDPG adopts a target actor network $\pi^{'}_{\phi^{'}}(s)$ and a target critic network $Q^{'}_{\theta^{'}}\left(s, a\right)$ to calculate the TD target value $y^{(t)}$ by
\begin{equation}
y^{(t)}=r^{(t)}+\zeta Q^{'}_{\theta^{'}}(s^{(t+1)}, \pi^{'}_{\phi^{'}}(s^{(t+1)}))\tag{24}.
\end{equation}
To minimize the loss function $L(\theta)$, parameter $\theta$ can be updated by the gradient
descent method with
\begin{equation}
\theta\leftarrow\theta-\alpha_{\theta}\nabla_{\theta}L\left(\theta\right)\tag{25},
\end{equation}
where $\alpha_{\theta}$ denotes the learning rate of the critic network. The parameters of the target networks, $\phi^{'}$ and $\theta^{'}$, are updated by the
Polyak averaging method \cite{712192} with
\begin{equation}
\phi^{'}\leftarrow\delta\phi+(1-\delta)\phi^{'}\tag{26},
\end{equation}
and
\begin{equation}
\theta^{'}\leftarrow\delta\theta+(1-\delta)\theta^{'}\tag{27},
\end{equation}
where $\delta\in\left[0, 1\right]$ denotes the soft update rate.

As DDPG is a deterministic policy algorithm which may get stuck in local optima, to efficiently explore the solution space, we add noise $n_{a}$ to the action during the training process with $n_{a}$ assumed to be a $\left|{\mathcal{A}}\right|$-dimensional sequence of entries modeled as i.i.d Gaussian random variables with zero mean and variance $\sigma_{a}^{2}$. To balance exploration and exploitation, we linearly decay the standard deviation $\sigma_{a}$ over training episode $e$ as
\begin{equation}
\sigma_{a}=\mathrm{max}\left\{\sigma_{a, max} - \kappa\cdot\hspace{0.5mm}e, \sigma_{a, min}\right\}\tag{28},
\end{equation}
where $\sigma_{a, max}$ and $\sigma_{a, min}$ denote the maximum and the minimum standard deviation, respectively. $\kappa$ represents the decay rate. In addition, to improve the efficiency of data utilization, a replay memory with capacity $D$ is adopted to store interaction experiences $\left(s^{(t)}, a^{(t)}, r^{(t)}, s^{(t+1)}\right)$, and a mini-batch of experiences with size $B$ are randomly sampled from the replay memory in each training step. The complete training process of our DDPG-C$^{2}$F framework is summarized in Algorithm 1.

\subsection{Complexity Analysis of Proposed DDPG-C$^{2}$F}
The computational complexity of the proposed DDPG-C$^{2}$F framework during the inference process is determined by the forward propagation of its actor network. 
Let $N_{A}$ denote the number of hidden layers in the actor network and $N_{a}^{n}$ denote the number of neurons in its $n$th layer. Since the actor network maps the state to an action, the number of neurons in the input layer is $L+\left|{\mathcal{A}}\right|$ and the number of neurons in the output layer is $\left|{\mathcal{A}}\right|$. The computational complexity of the proposed DDPG-C$^{2}$F framework is therefore $O((L\hspace{-0.5mm}+\hspace{-0.5mm}\left|{\mathcal{A}}\right|)\hspace{-0.5mm}\times\hspace{-0.5mm}N_{a}^{1}\hspace{-0.5mm}+\hspace{-0.5mm}\sum_{n=1}^{N_{A}-1}\hspace{-0.75mm}N_{a}^{n}\hspace{-0.25mm}\times\hspace{-0.25mm}N_{a}^{n+1}\hspace{-0.5mm}+\hspace{-0.25mm}N_{a}^{N_{A}}\hspace{-0.5mm}\times\hspace{-0.25mm} \left|{\mathcal{A}}\right|)\hspace{-0.5mm}=\hspace{-0.5mm}O(L+\left|{\mathcal{A}}\right|)$.

\section{Case Studies: Rate-Oriented Problems}
Recall that the clustered cell-free networking problem is formulated as a general problem $\mathcal{P}1$, which can be tailored in different cases to accommodate different practical requirements. In this section, we apply the proposed DDPG-C$^{2}$F framework to two representative rate-oriented clustered cell-free networking problems to demonstrate its effectiveness.

\subsection{Rate-Oriented Problems}
Here, we consider the sum rate of the clustered cell-free network as the performance metric, expressed as
\begin{equation}
R_{sum}=\sum_{k=1}^{K} R_{k}\tag{29},
\end{equation}
where $R_{k}$ is the achievable rate of user $u_{k}$, and consider two representative rate-oriented problems.

\subsubsection{Joint Optimization of Sum Rate and Balance of Subnetworks} 
With the aim of maximizing both the sum rate and the balance of subnetworks, we consider the below objective function
\begin{equation}
f\left(\mathcal{C}^{(t)}\right)=R_{\rho}^{(t)}\triangleq R_{sum}^{(t)}\cdot\hspace{1mm}\rho^{(t)}\tag{30},
\end{equation}
where $R_{\rho}^{(t)}$ is referred to as balance-aware sum rate at time interval $t$. For the spatial precoding adopted in each subnetwork, there might be further requirements. For instance, zero-forcing beamforming (ZFBF) requires that the number of APs in a subnetwork is no less than the number of users in it. Such constraints can be reflected in
the constraint set $\mathcal{V}$ as
\begin{equation}
\mathcal{V}=\left\{\mathcal{C}^{(t)}:\mathcal{V}_{pre}\right\}\tag{31}.
\end{equation} 

\subsubsection{Maximization of Balance of Subnetworks with Sum Rate Constraint} 
In the scenario with certain sum rate requirement, the clustered cell-free networking problem can be formulated to maximize the balance of subnetworks under the constraint that the sum rate $R_{sum}$ must be no smaller than a given threshold $R_{th}$. In this case, the objective function is given by
\begin{equation}
f\left(\mathcal{C}^{(t)}\right)=\rho^{(t)}\tag{32},
\end{equation}
and the constraint set is
\begin{equation}
\mathcal{V}=\left\{\mathcal{C}^{(t)}:R_{sum}^{(t)}\geq R_{th}; \mathcal{V}_{pre}\right\}\tag{33}.
\end{equation}

Though with different objectives and constraints in these two cases, the users and APs close to each other should be assigned into the same subnetwork to improve the desired signal power and mitigate the mutual interference. Motivated by this, the feature of the anchor of the $m$th subnetwork at time interval $t$ can be defined by its location as
\begin{equation}
\boldsymbol{o}_{m}^{(t)}=\left(o_{m,x}^{(t)}, o_{m,y}^{(t)}\right)\tag{34},
\end{equation}
where $o_{m,x}^{(t)}$ and $o_{m,y}^{(t)}$ are the coordinates of the anchor. Users and APs can then be affiliated with their closest subnetwork anchors to form subnetworks. In this setting, the dimensionality of the action space is $\left|{\mathcal{A}}\right|=2M$.

\subsection{Simulation Results}
In the simulations, we consider a network with users and APs randomly distributed in a 1000 m $\times$ 1000 m square area, and employ the random walk (RW) mobility model \cite{8673556} to simulate user movements. Specifically, each user moves with a random speed $V\in[0, V_{max}]$ in a direction randomly selected from the range of $[0, 2\pi]$ in each time interval. The time horizon includes $T=100$ time intervals and each interval is of one second.

The large-scale fading coefficient from AP $b_{l}$ to user $u_{k}$ is modeled as
\begin{equation}
\gamma_{k, l}=\sqrt{d_{k, l}^{-\alpha}\cdot10^{\frac{\chi_{k, l}}{10}}}\tag{35},
\end{equation}
where $d_{k, l}$ represents the distance between user $u_{k}$ and AP $b_{l}$, and $\alpha$ is the path-loss exponent. $\chi_{k, l} \sim \mathcal{N}(0, \sigma_{sh}^{2})$ denotes the shadowing factor with zero mean and standard deviation $\sigma_{sh}$. Note that although the channel feature in the state space accounts for large-scale fading only, small-scale fading is considered for performance evaluation. The average transmission power of each AP is assumed to be fixed at $P$ and the total transmission power of the APs in a subnetwork is equally allocated to the users therein. In each subnetwork, ZFBF is adopted, and the required $\mathcal{V}_{pre}$ is specified as
\begin{equation}
\mathcal{V}_{pre}=\left\{L_{m}^{(t)}\geq K_{m}^{(t)},~\forall m\right\}\tag{36}.
\end{equation}

\begin{table}[!t]
\caption{Simulation parameters}
\centering
\renewcommand{\arraystretch}{1.1}
\begin{tabular}{|l|l|}
\hline
Parameter & Value \\
\hline
Noise power $N_{0}$ & -104dBm \\
Path-loss exponent $\alpha$ & 4  \\
Shadow fading standard deviation $\sigma_{sh}$ & 8dB \\
Transmission power per-AP $P$ & 2W \\
Circuit power consumption $P_{c}$ & 1W \\
Fixed power consumption $P_{fix}$ & 0.05W \\
Traffic-dependent power consumption $P_{b}$ & 0.1W/bps/Hz \\
Power amplifier efficiency $\tau$ & 38\% \\
Replay memory capacity $D$ & 10000 \\
Mini-batch size $B$ & 128 \\
Actor network learning rate $\alpha_{\phi}$ & 0.0001 \\
Critic network learning rate $\alpha_{\theta}$ & 0.001  \\
Soft update rate $\delta$ & 0.001 \\
Discount factor $\zeta$ & 0.99 \\
Action noise standard deviations $\sigma_{a, max}, \sigma_{a, min}$ & 0.25, 0.001 \\
Action noise decay rate $\kappa$ & 0.0001 \\
\hline
\end{tabular}
\end{table}

For the proposed DDPG-C$^{2}$F framework, we employ two
hidden layers in the actor network with 256 and 128 neurons, respectively. The critic network has three hidden layers with 512, 256 and 128 neurons, respectively. In addition, ReLU is adopted as the activation function and Adam optimizer is applied to update the neural network parameters. We train 4000 episodes for the proposed DDPG-C$^{2}$F framework, each with 100 steps. Other essential parameters are summarized in Table I, which are used throughout this paper. All simulations are conducted on a computer with an Intel Core i7-12700K CPU and an NVIDIA GeForce RTX 3050 graphics processing unit (GPU).

To demonstrate the performance of the proposed DDPG-C$^{2}$F framework, we compare our framework with three existing representative clustered cell-free networking algorithms:
\begin{itemize}
\item \textbf{AP-centric clustering}\cite{10535986}: The AP-centric benchmark in \cite{10535986} first groups APs with GMM algorithm,  and then assigns each user to the closest AP cluster.
\item \textbf{User-centric clustering}\cite{8644255}: The user-centric benchmark in \cite{8644255} first clusters users with K-means algorithm, and then assign each AP to the closest user cluster.
\item \textbf{Graph partitioning}\cite{JunyuanTWC}: The graph partitioning benchmark in \cite{JunyuanTWC} first merges each user and its closest AP as a meganode, and then applies spectral clustering algorithm to cluster meganodes for network partitioning.
\end{itemize}

\subsubsection{Joint Optimization of Sum Rate and Balance of Subnetworks} 
Fig. 3 plots the balance-aware sum rate $R_{\rho}$ achieved during the training process with the proposed DDPG-C$^{2}$F framework and that achieved by the benchmarks. To validate the effectiveness of selecting the large-scale fading coefficient of the strongest channel between each AP and all users as its feature in the state space, the result achieved by a variant of the proposed framework that utilizes all large-scale fading coefficients between all APs and all users is also plotted for comparison. As shown in Fig. 3, the balance-aware sum rates $R_{\rho}$ with both the proposed DDPG-C$^{2}$F framework and its variant first increase and then converge at the same value after 2500 episodes, corroborating that considering the strongest channel between each AP and the users is sufficient for training the proposed DDPG-C$^{2}$F framework. That is, our DDPG-C$^{2}$F framework can significantly reduce the channel estimation costs and computational complexity without bringing performance degradation. Moreover, we can clearly see that the proposed DDPG-C$^{2}$F framework achieves much higher balance-aware sum rate than other baselines.

\begin{figure}[t]
\centering                             
\includegraphics[width=0.42\textwidth]{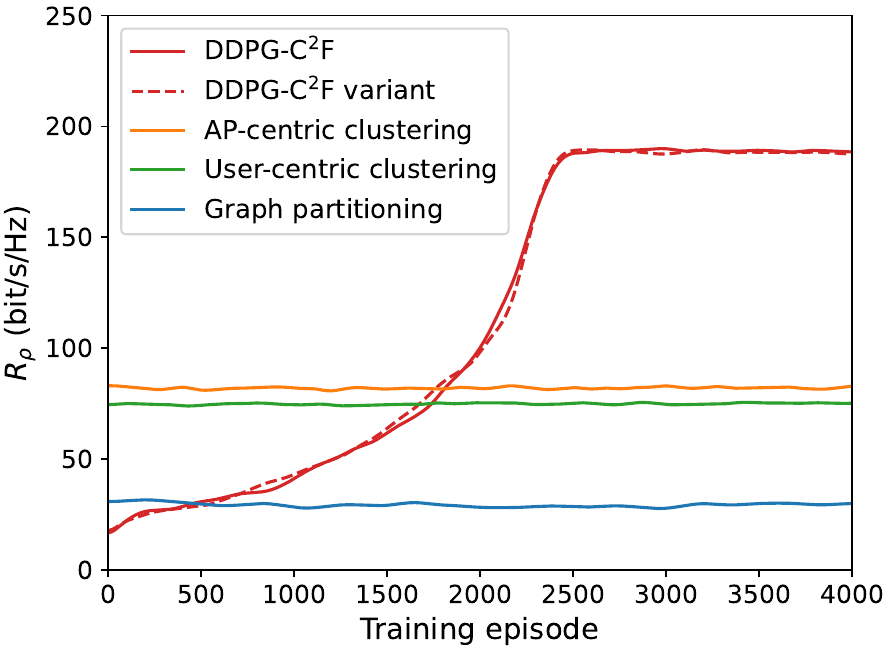}\\
\caption{Balance-aware sum rate $R_{\rho}$ during the training process in the case of joint optimization of sum rate and balance of subnetworks. $K=50$. $L=100$. $M=5$. $V_{max}=5$m/s.}
\end{figure}
\begin{figure*}[t]
\centering
\subfloat[AP-centric clustering]{\includegraphics[width=0.24\textwidth]{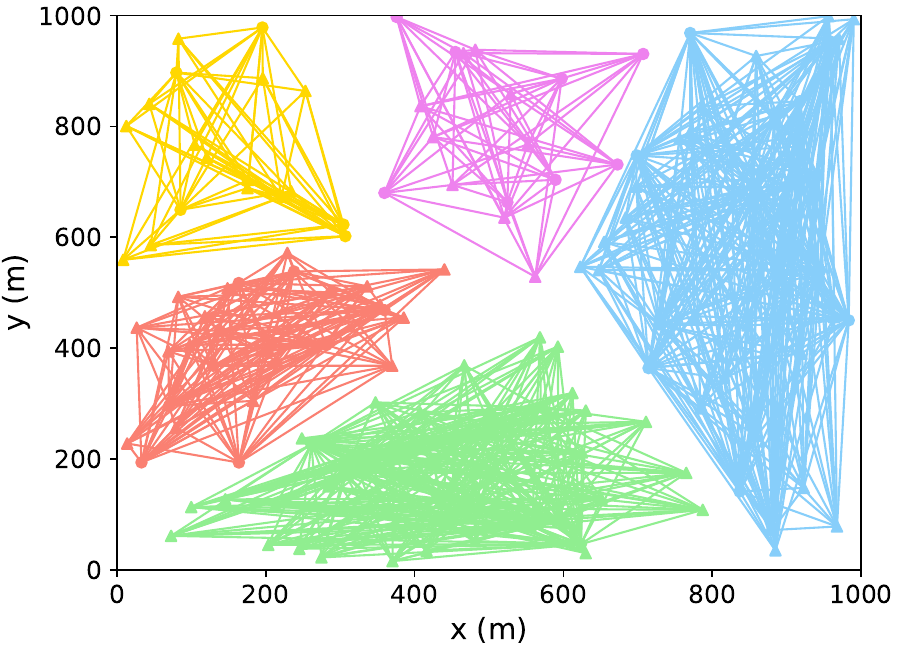}}
\hspace{0.5mm}
\subfloat[User-centric clustering]{\includegraphics[width=0.24\textwidth]{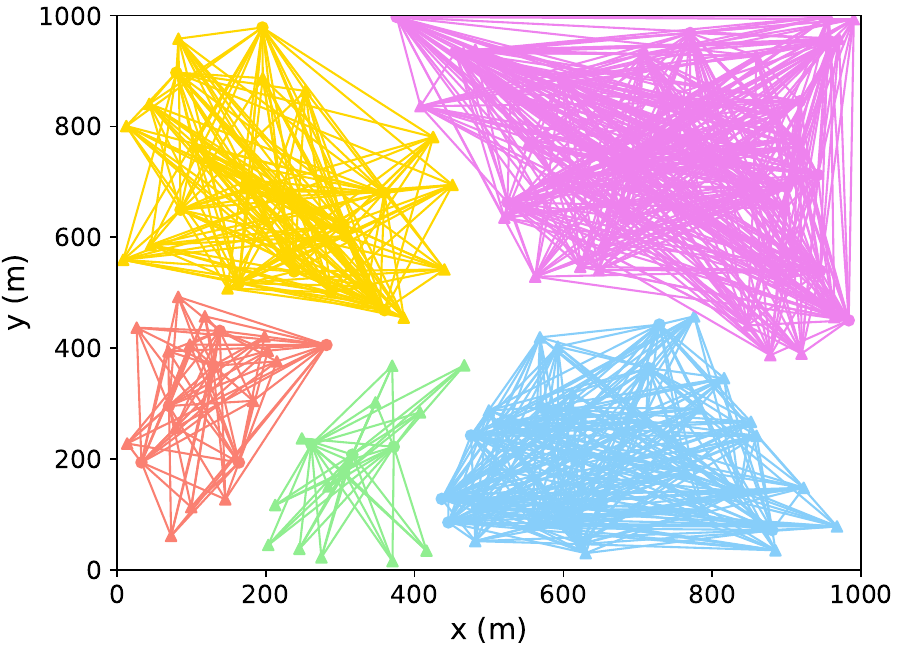}}
\hspace{0.5mm}
\subfloat[Graph partitioning]{\includegraphics[width=0.24\textwidth]{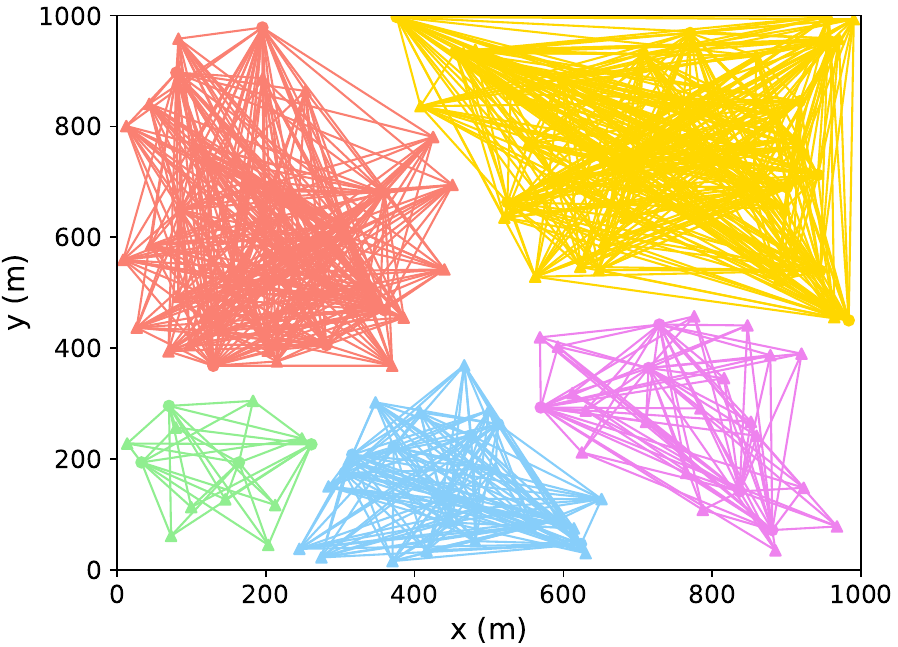}}
\hspace{0.5mm}
\subfloat[DDPG-C$^{2}$F]{\includegraphics[width=0.24\textwidth]{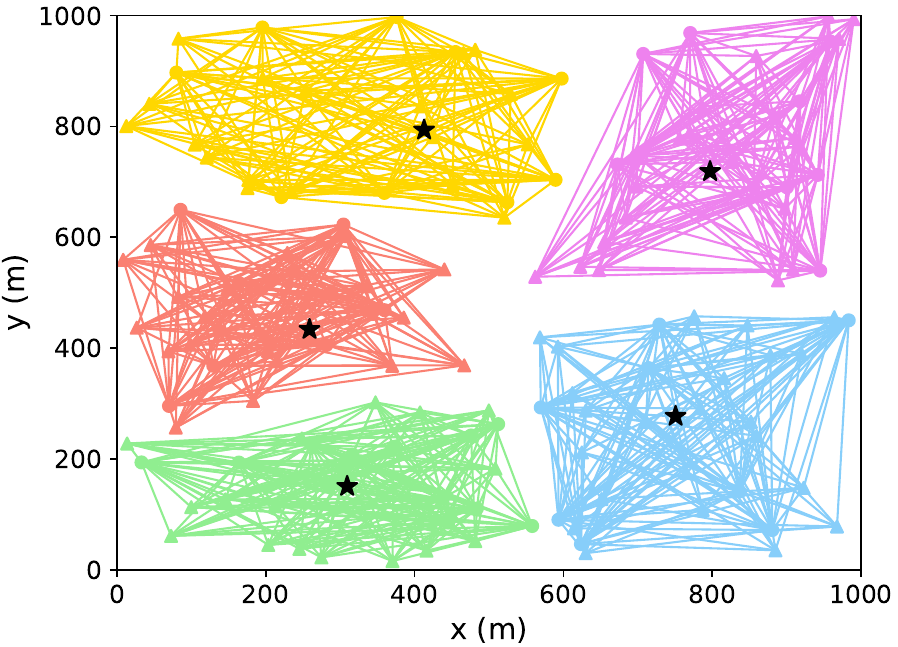}}
\caption{Clustered cell-free networking results of a random network snapshot with the proposed DDPG-C$^{2}$F framework and the benchmarks in the case of joint optimization of sum rate and balance of subnetworks. ``$\circ$'' represents a user and ``$\triangle$'' represents an AP. ``$\star$'' represents the subnetwork anchor. Users and APs in the same subnetwork are connected by lines in the same color. $K=50$. $L=100$. $M=5$. $V_{max}=5$m/s.}
\end{figure*}

\begin{figure*}[t]
\centering
\subfloat[$K=\{25, 50, 75\}$]{\includegraphics[width=0.24\textwidth]{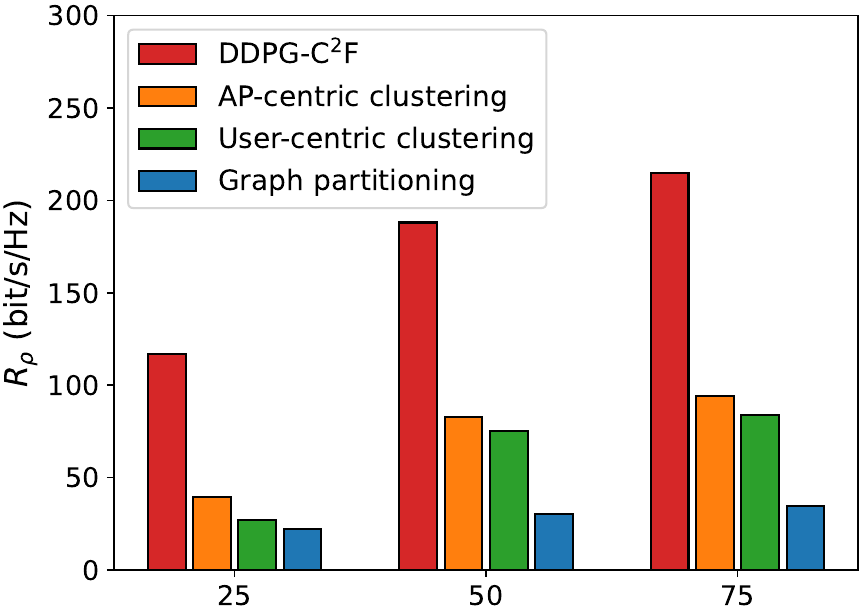}}
\hspace{0.5mm}
\subfloat[$L=\{75, 100, 200\}$]{\includegraphics[width=0.24\textwidth]{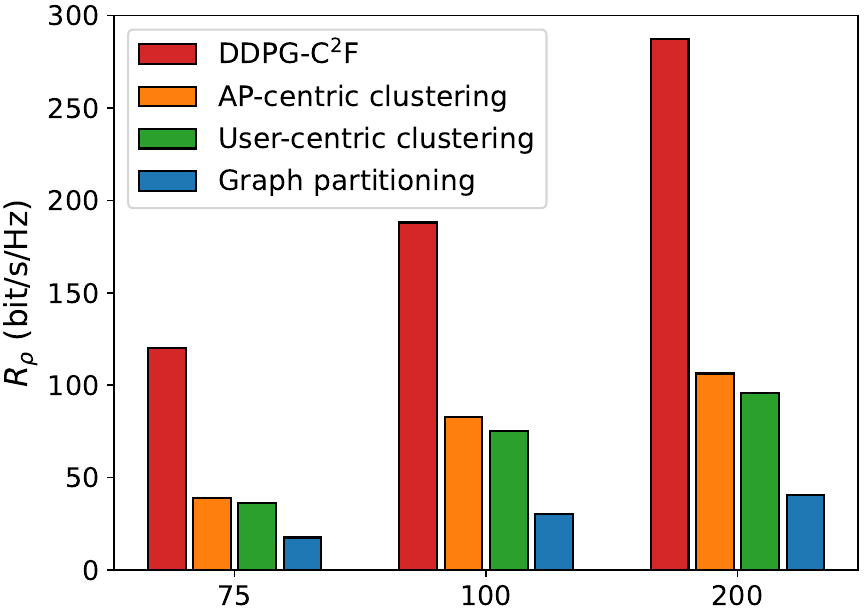}}
\hspace{0.5mm}
\subfloat[$M=\{3, 5, 7\}$]{\includegraphics[width=0.24\textwidth]{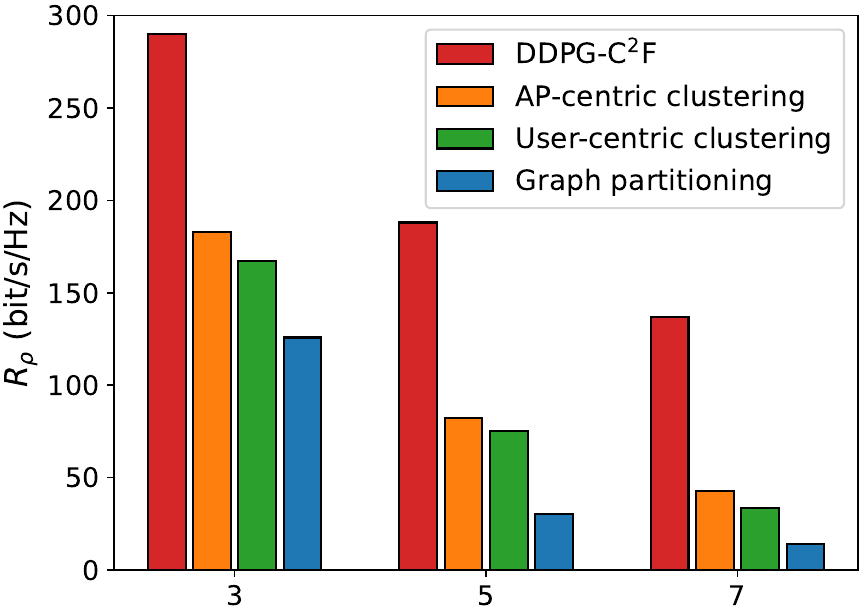}}
\hspace{0.5mm}
\subfloat[$V_{max}=\{1, 5, 10\}$m/s]{\includegraphics[width=0.24\textwidth]{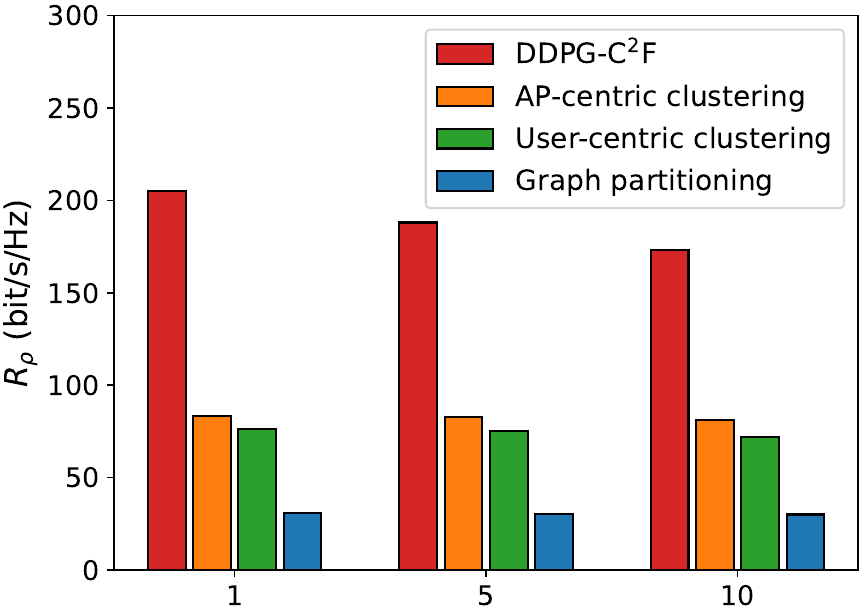}}
\caption{Balance-aware sum rate $R_{\rho}$ with the proposed DDPG-C$^{2}$F framework and the benchmarks in the case of joint optimization of sum rate and balance of subnetworks. By default, $K=50$, $L=100$, $M=5$ and $V_{max}=5$m/s.}
\end{figure*}

\begin{figure*}[t]
\centering
\subfloat[]{
\label{1a}
\includegraphics[width=0.30\textwidth]{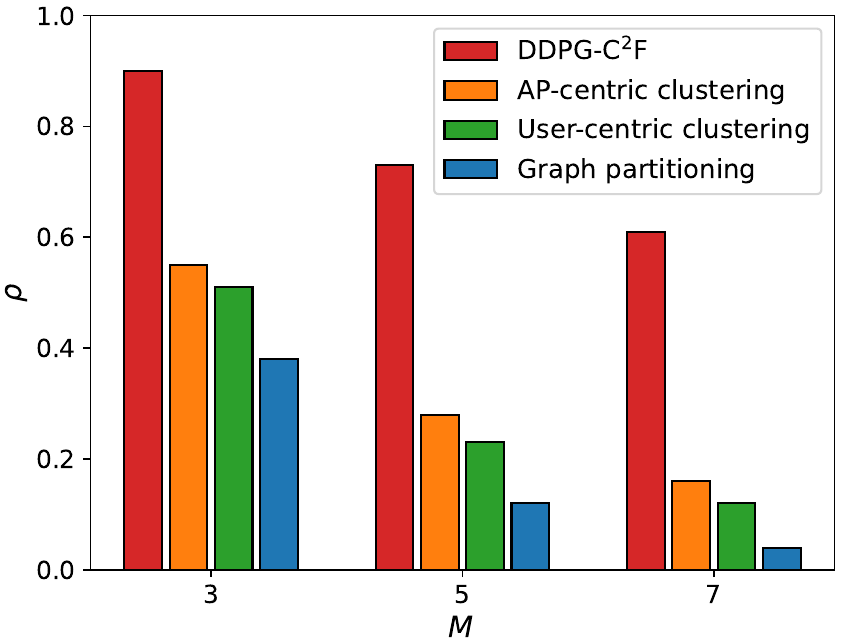}}
\hspace{1mm}
\subfloat[]{
\label{1b}
\includegraphics[width=0.30\textwidth]{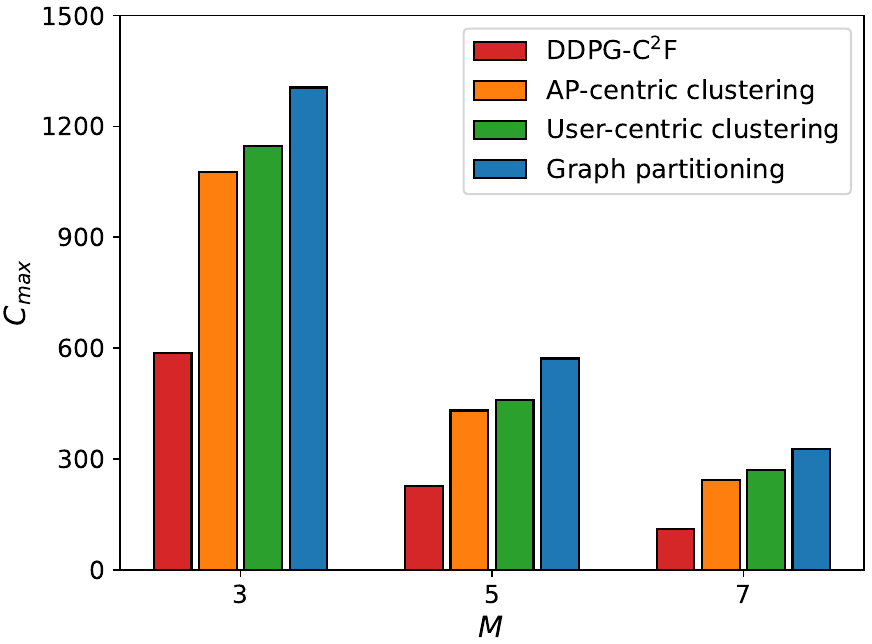}}
\hspace{1mm}
\subfloat[]{
\label{1c}
\includegraphics[width=0.30\textwidth]{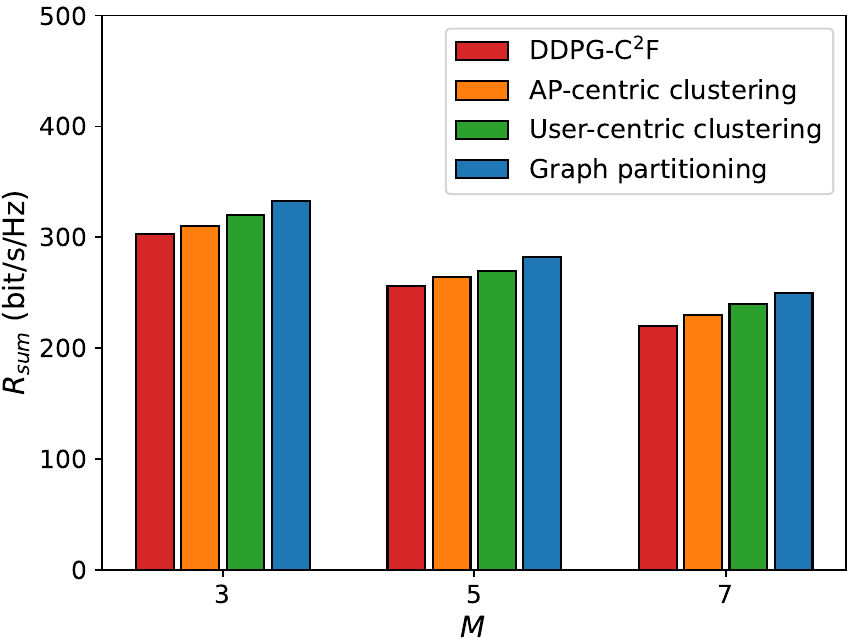}}
\caption{(a) Balance of subnetworks $\rho$, (b) maximum number of channels across all subnetworks $C_{max}$ and (c) sum rate $R_{sum}$ versus the number of subnetworks $M$ in the case of joint optimization of sum rate and balance of subnetworks. $K=50$. $L=100$. $V_{max}=5$m/s.}
\end{figure*}

\begin{figure*}[t]
\centering
\subfloat[$R_{th}=250$ bit/s/Hz]{
\label{1a}
\includegraphics[width=0.30\textwidth]{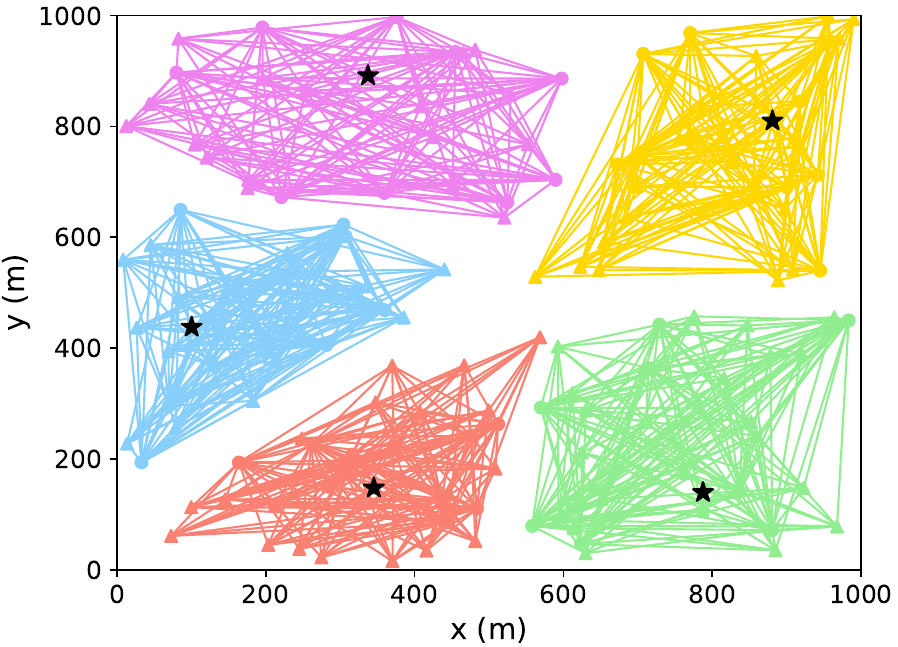}}
\hspace{1mm}
\subfloat[$R_{th}=300$ bit/s/Hz]{
\label{1b}
\includegraphics[width=0.30\textwidth]{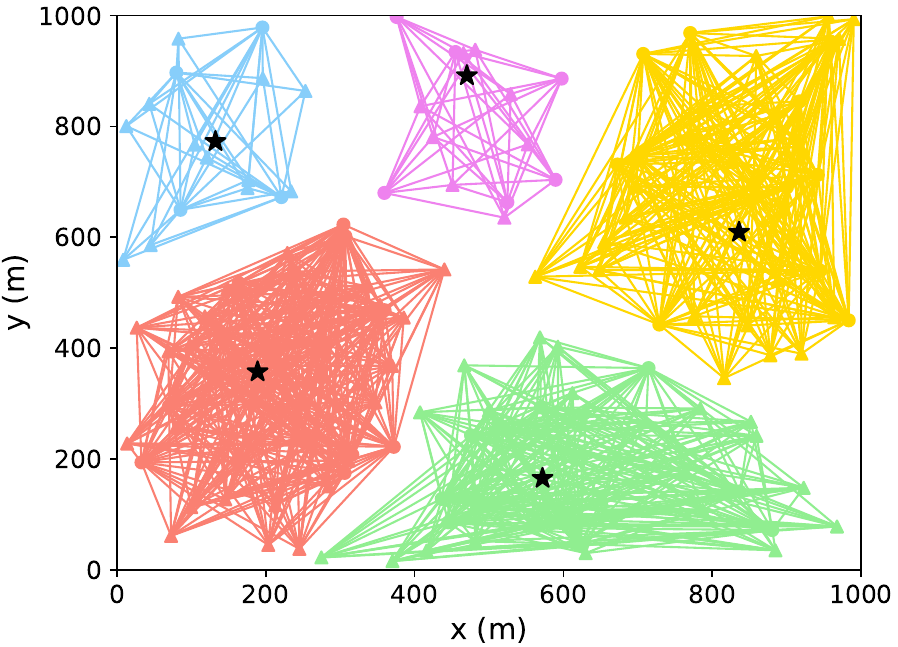}}
\hspace{1mm}
\subfloat[$R_{th}=350$ bit/s/Hz]{
\label{1c}
\includegraphics[width=0.30\textwidth]{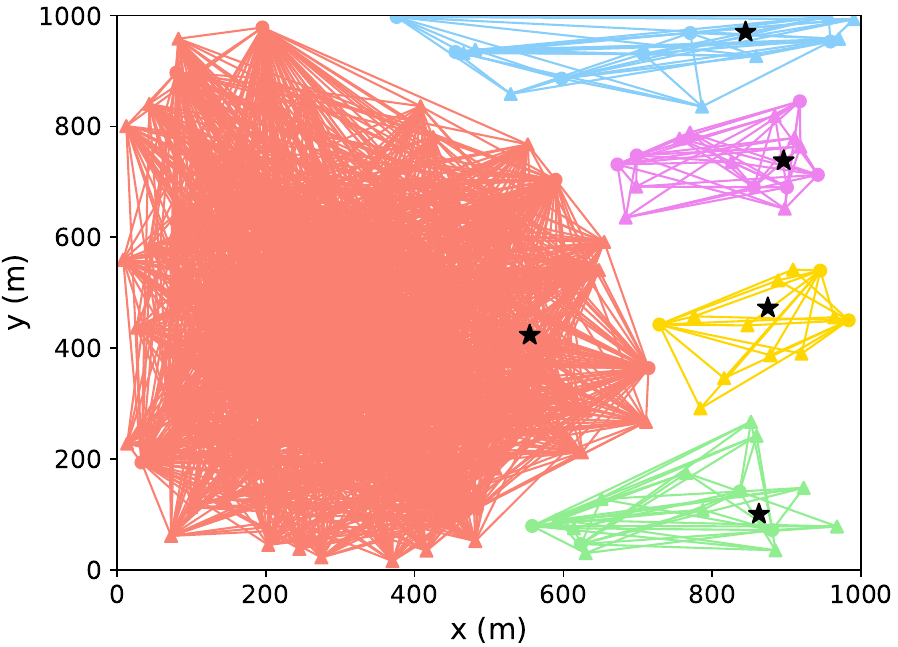}}
\caption{Clustered cell-free networking results of a random network snapshot with the proposed DDPG-C$^{2}$F framework under various rate threshold $R_{th}$ in the case of maximization of balance of subnetworks with sum rate constraint. ``$\circ$'' represents a user and ``$\triangle$'' represents an AP. ``$\star$'' represents the subnetwork anchor. Users and APs in the same subnetwork are connected by lines in the same color. $K=50$. $L=100$. $M=5$. $V_{max}=5$m/s.}
\end{figure*}

Fig. 4 illustrates the clustered cell-free networking results of a random network snapshot with the proposed DDPG-C$^{2}$F framework and the benchmarks. It can be observed from Figs. 4(a)--4(d) that compared to the benchmarks, the proposed DDPG-C$^{2}$F framework produces more balanced subnetworks, corroborating its superior performance in efficiently reducing the signaling overhead and computational complexity of joint transmission in each subnetwork.

Fig. 5 further presents the balance-aware sum rate $R_{\rho}$ with 
the proposed DDPG-C$^{2}$F framework and the benchmarks by varying the number of users $K$, the number of APs $L$, the number of subnetworks $M$, and the maximum speed of users $V_{max}$, respectively. We can see from Fig. 5 that the proposed framework achieves the highest balance-aware sum rate in all cases. Particularly, Fig. 5(b) shows that more prominent gains can be achieved as the number of APs $L$ increases. This is because the proposed framework can always balance the numbers of APs across subnetworks, while the imbalance issue of the benchmarks could be enlarged when the number of APs is large. Additionally, it is shown in Fig. 5(d) that the balance-aware sum rate $R_{\rho}$ achieved by the proposed DDPG-C$^{2}$F framework with $V_{max}=10$m/s is just slightly lower than that with $V_{max}=1$m/s, indicating that the proposed framework can maintain its superior performance in high-mobility scenarios.

To have a closer look at the balance of subnetworks and the rate performance, Fig. 6 presents the balance of subnetworks $\rho$, the maximum number of channels across all subnetworks $C_{max}=\mathrm{max}_{m=1,2, \cdots, M}K_{m}L_{m}$,\footnote{Note that instead of measuring the size of a subnetwork by the total number of users and APs in it, we are more interested in the product of the number of users and the number of APs, i.e., the number of channels in it, because the number of channels determines the channel measurement costs for implementing spatial precoding.} and the sum rate $R_{sum}$. By comparing Fig. 6(a) and Fig. 6(b), it can be found that the proposed DDPG-C$^{2}$F framework leads to much higher balance of subnetworks $\rho$ and much lower $C_{max}$, corroborating that the subnetwork sizes could be effectively balanced by maximizing the defined balance metric $\rho$. In addition, Fig. 6(c) shows that the proposed DDPG-C$^{2}$F framework achieves similar sum rate performance as the benchmarks. The above results indicate that the proposed framework can significantly reduce the signaling overhead and joint processing complexity while maintaining similar rate performance.

\subsubsection{Maximization of Balance of Subnetworks with Sum Rate Constraint} 
Fig. 7 illustrates the clustered cell-free networking results with the proposed DDPG-C$^{2}$F framework under various sum rate constraints. It is shown in Fig. 7(a) that with a relatively low threshold $R_{th}=250$ bit/s/Hz, the whole network is partitioned into balanced subnetworks with similar subnetwork sizes. By increasing the rate threshold to $R_{th}=350$ bit/s/Hz, a giant subnetwork appears. Intuitively, when a sufficiently high data rate is required, more and more users and APs would be grouped into the same subnetwork to increase the joint processing gain.

\begin{figure}[t]
\centering                             
\includegraphics[width=0.42\textwidth]{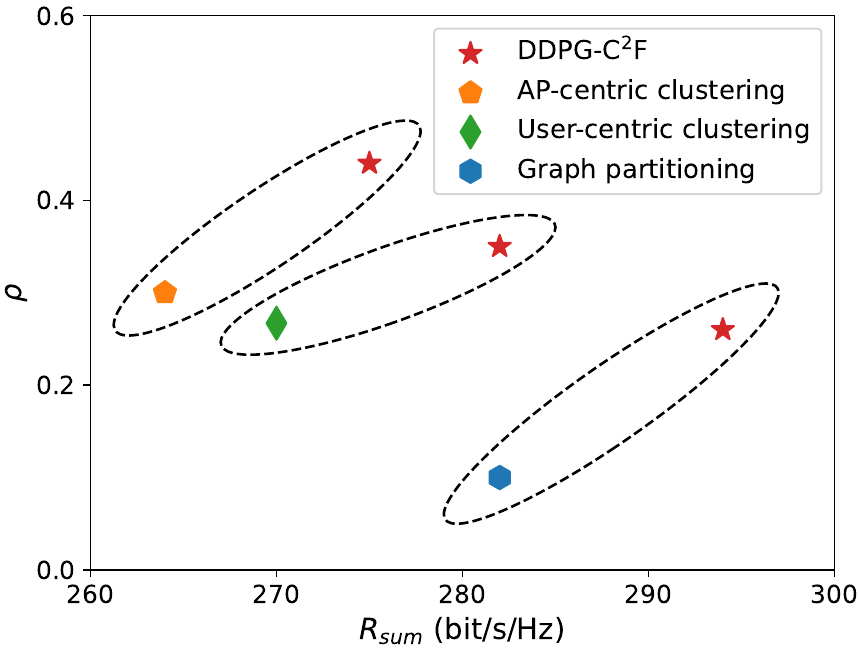}\\
\caption{Balance of subnetworks $\rho$ and sum rate $R_{sum}$ with the proposed DDPG-C$^{2}$F framework and the benchmarks in the case of maximization of balance of subnetworks with sum rate constraint. The results of the proposed framework are obtained by setting the rate threshold $R_{th}$ as the sum rate achieved with the benchmarks in the same dashed circle. $K=50$. $L=100$. $M=5$. $V_{max}=5$m/s.}
\end{figure}

Fig. 8 compares the proposed DDPG-C$^{2}$F framework and the benchmarks in terms of balance of subnetworks $\rho$ and sum rate $R_{sum}$, where the rate threshold $R_{th}$ in the proposed DDPG-C$^{2}$F framework is set the same as the sum rate achieved with the benchmark in the same dashed circle. We can observe from Fig. 8 that the proposed DDPG-C$^{2}$F framework meets the sum rate constraint, and notably, achieves higher sum rate and higher balance of subnetworks than the benchmarks.

\subsubsection{Comparison of Computational Complexity and Running Time} 
Table II compares the computational complexity and running time of the proposed DDPG-C$^{2}$F framework with the benchmarks. As mentioned in Section IV-A, the dimensionality of the action space is $\left|{\mathcal{A}}\right|=2M$. Consequently, according to Section III-D, the proposed DDPG-C$^{2}$F framework has the computational complexity $O(L+2M)=O(L)$, since the number of subnetworks $M$ is usually much smaller than the number of APs $L$. The computational complexities of the AP-centric clustering and the user-centric clustering were given in [14] and [15] as $O(K^{3}L)$ and $O(KL)$, respectively, while the graph partitioning benchmark has a higher complexity of $O(L^{3})$ [12]. We can clearly see from Table II that the proposed DDPG-C$^{2}$F framework incurs much lower computational cost and running time than the benchmarks.

\begin{table}[t]
\centering
\caption{Comparison of Computational Complexity and Running Time in Rate-Oriented Cases. $M=5$, $K=50$, $L=100$.}
\begin{tabular}{|c|c|c|} \hline
\textbf{Algorithms} & \textbf{Computational complexity} & \textbf{Running time (s)} \\ \hline
DDPG-C$^{2}$F & $O(L)$ & 0.07 \\ \hline
AP-centric clustering & $O(K^{3}L)$ & 1.21 \\ \hline
User-centric clustering & $O(KL)$ & 0.11 \\ \hline
Graph partitioning & $O(L^{3})$ & 11.04 \\ \hline
\end{tabular}
\end{table}

\section{Case Studies: Energy Efficiency-Oriented Problems}
In this section, problem $\mathcal{P}1$ is tailored for energy efficiency optimization, which is another major concern in practical wireless communication systems due to the huge power consumption resulting from densely deployed APs. Two representative energy efficiency-oriented clustered cell-free networking problems are studied by applying the proposed DDPG-C$^{2}$F framework.

\subsection{Energy Efficiency-Oriented Problems}
The energy efficiency $\eta_{ee}$ of the clustered cell-free network is defined as
\begin{equation}
\eta_{ee}=\frac{R_{sum}}{P_{tot}}\tag{37},
\end{equation}
where $P_{tot}$ denotes the total power consumption, consisting of the power consumed by the APs, $P_{A}$, and the fronthaul power consumption $P_{B}$, i.e.,
\begin{equation}
P_{tot}=P_{A}+P_{B}\tag{38}.
\end{equation}
The power consumption of the APs in all the $M$ subnetworks can be modeled as \cite{7062017}
\begin{equation}
P_{A}=\sum_{m=1}^{M}\left(\frac{PL_{m}}{\tau}+P_{c}L_{m}\right)\tag{39},
\end{equation}
where $\tau$ represents the power amplifier efficiency, and $P_{c}$ is the circuit power consumed by frequency synthesizers, mixers and transmit filters, etc. The power consumption of
the fronthaul links, $P_{B}$, can be written as \cite{9148948}
\begin{equation}
P_{B}=P_{\emph{fix}}L+P_{b} R_{sum}\tag{40},
\end{equation}
where $P_{\emph{fix}}$ is the power consumed regardless of the fronthaul traffic, and $P_{b}$ denotes the traffic-dependent power consumption. By substituting (39) and (40) into (38), the total power consumption $P_{tot}$ can be obtained as
\begin{equation}
P_{tot}=\left(\frac{P}{\tau}+P_{c}\right)\sum_{m=1}^{M}L_{m}+P_{\emph{fix}}L+P_{b} R_{sum}\tag{41}.
\end{equation}
We then introduce the two energy efficiency-oriented problems.

\subsubsection{Joint Optimization of Energy Efficiency and Balance of Subnetworks} 
Similar to the rate-oriented case, the objective function of simultaneously maximizing energy efficiency and the balance of subnetworks is defined as
\begin{equation}
f\left(\mathcal{C}^{(t)}\right)=\eta_{\rho}^{(t)}\triangleq\eta_{ee}^{(t)}\cdot\hspace{1mm}\rho^{(t)}\tag{42},
\end{equation}
where $\eta_{\rho}^{(t)}$ is the balance-aware energy efficiency at time interval $t$. The constraint set $\mathcal{V}$ is the same as that given in (31).

\subsubsection{Joint Optimization of Energy Efficiency and Balance of Subnetworks with Sum Rate Constraint} Note that maximizing energy efficiency solely might result in very low sum rate and thus poor user experience or even service outage. Therefore, we further consider the scenario with some sum rate constraint. In this case, the objective function $f\left(\mathcal{C}^{(t)}\right)$ remains the same as that in (42), and the constraint is the same as that in (33).

\begin{figure}[t]
\centering                             
\includegraphics[width=0.42\textwidth]{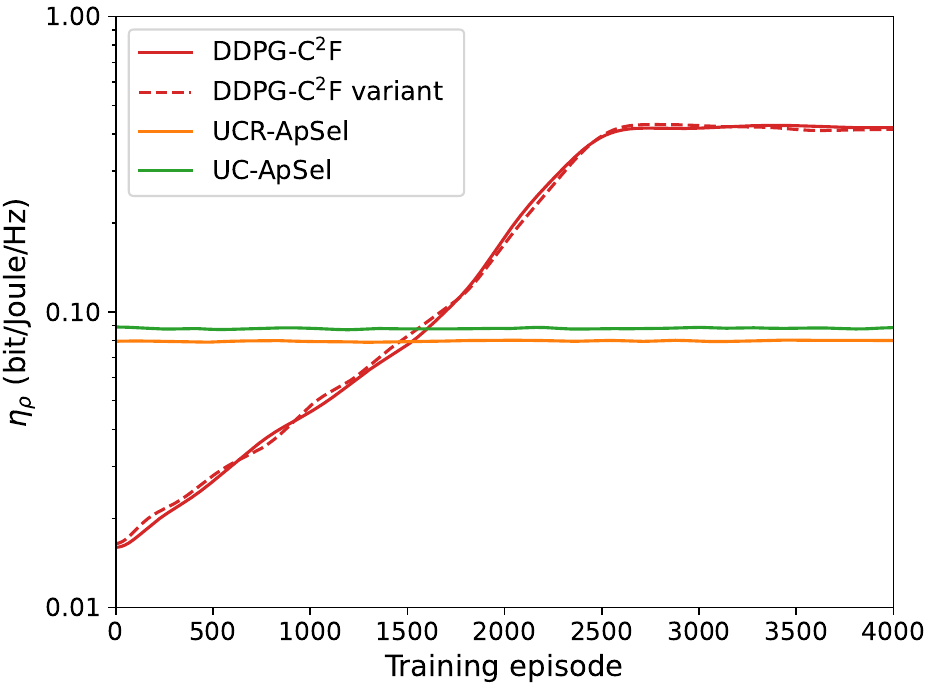}\\
\caption{Balance-aware energy efficiency $\eta_{ee}$ during the training process in the case of joint optimization of energy efficiency and balance of subnetworks. $K=50$. $L=100$. $M=5$. $V_{max}=5$m/s.}
\end{figure}

Intuitively, to maximize the energy efficiency, there is no need to involve the APs far away from users for data transmission, as these APs contribute little to the desired signal power yet consume loads of energy and cause interference to other subnetworks. To enable the dynamic on-off switch of APs in the proposed DDPG-C$^{2}$F framework, we include an AP-selection weight into the feature of each subnetwork anchor. The anchor feature of the $m$th subnetwork at time interval $t$ is now defined as
\begin{equation}
\boldsymbol{o}_{m}^{(t)}=\left(o_{m,x}^{(t)}, o_{m,y}^{(t)}, o_{m, w}^{(t)}\right)\tag{43},
\end{equation}
where the AP-selection weight $o_{m, w}^{(t)}\in\left(0, 1\right)$ is the ratio of the number of APs that should be activated to the total number of APs in the $m$th subnetwork. After learning the anchor features, users and APs are first affiliated with their closest anchors to form subnetworks, and then in each subnetwork, the APs close to users are selected for transmission according to the AP-selection weight. With this setting, the dimensionality of the action space is $\left|{\mathcal{A}}\right|=3M$.

\begin{figure*}[t]
\centering
\subfloat[$K=\{25, 50, 75\}$]{\includegraphics[width=0.24\textwidth]{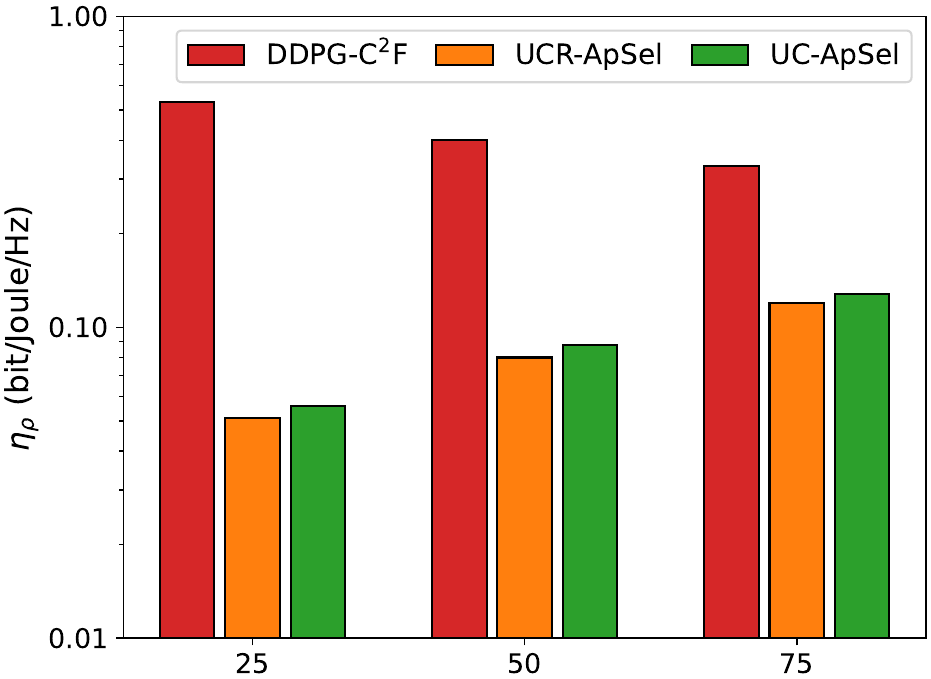}}
\hspace{0.5mm}
\subfloat[$L=\{75, 100, 200\}$]{\includegraphics[width=0.24\textwidth]{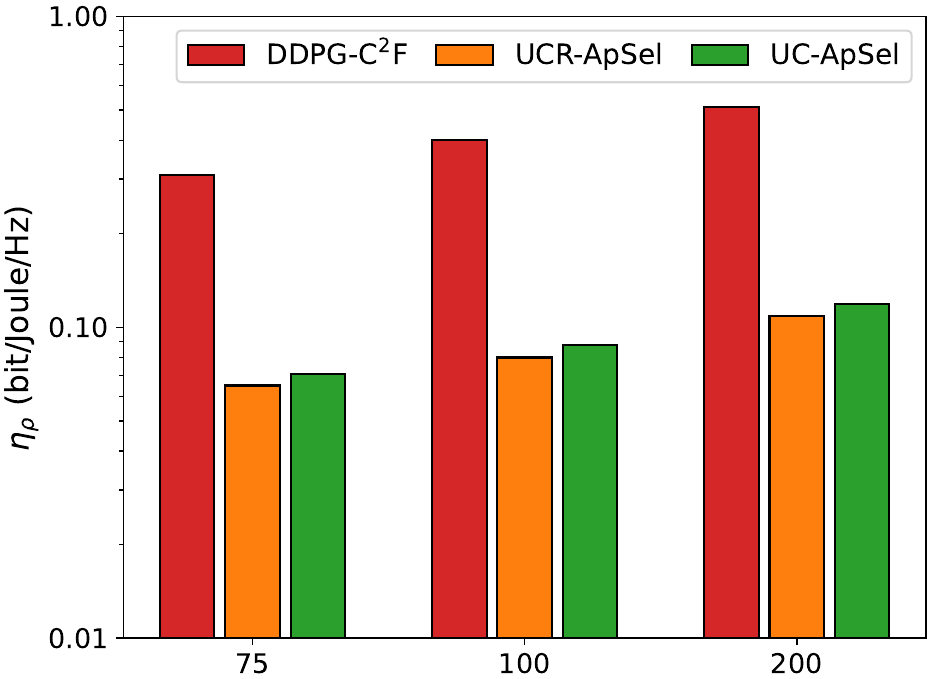}}
\hspace{0.5mm}
\subfloat[$M=\{3, 5, 7\}$]{\includegraphics[width=0.24\textwidth]{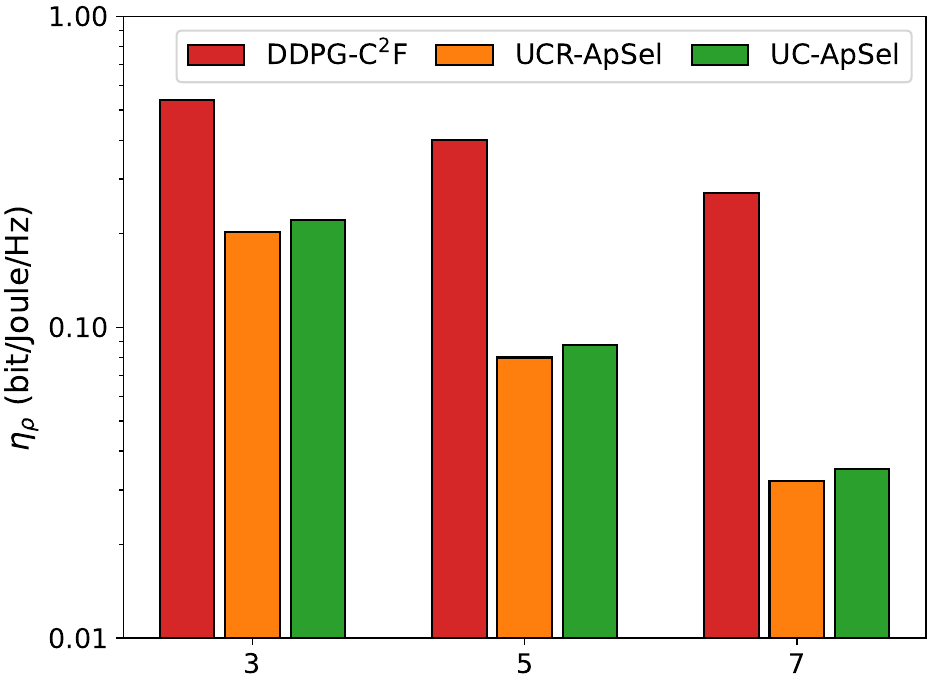}}
\hspace{0.5mm}
\subfloat[$V_{max}=\{1, 5, 10\}$m/s]{\includegraphics[width=0.24\textwidth]{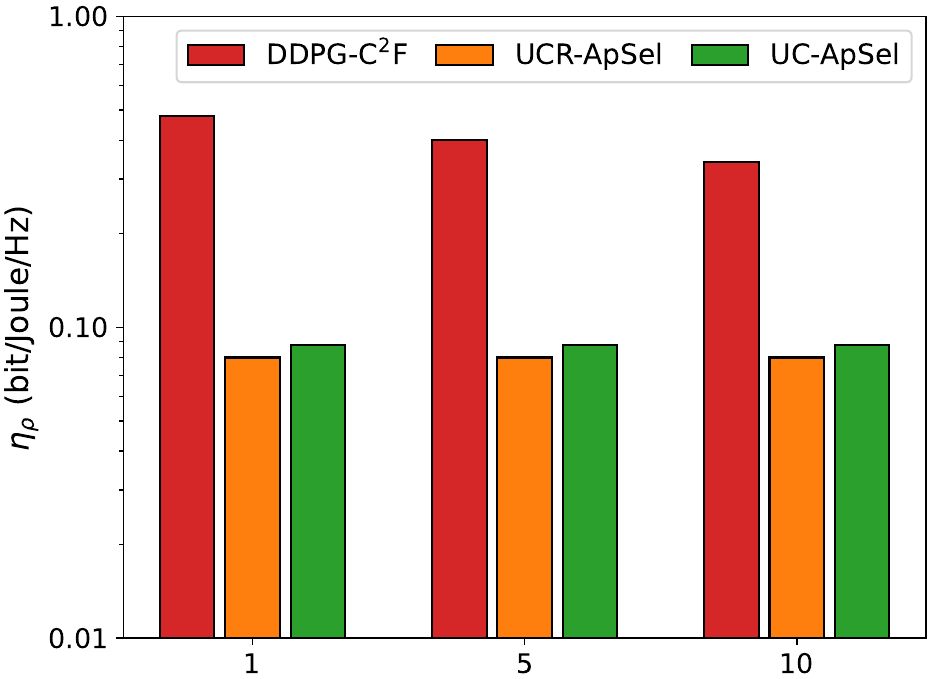}}
\caption{Balance-aware energy efficiency $\eta_{\rho}$ with the proposed DDPG-C$^{2}$F framework and the benchmarks in the case of joint optimization of energy efficiency and balance of subnetworks. By default, $K=50$, $L=100$, $M=5$ and $V_{max}=5$m/s.}
\end{figure*}

\begin{figure*}[t]
\centering
\subfloat[]{
\label{1a}
\includegraphics[width=0.30\textwidth]{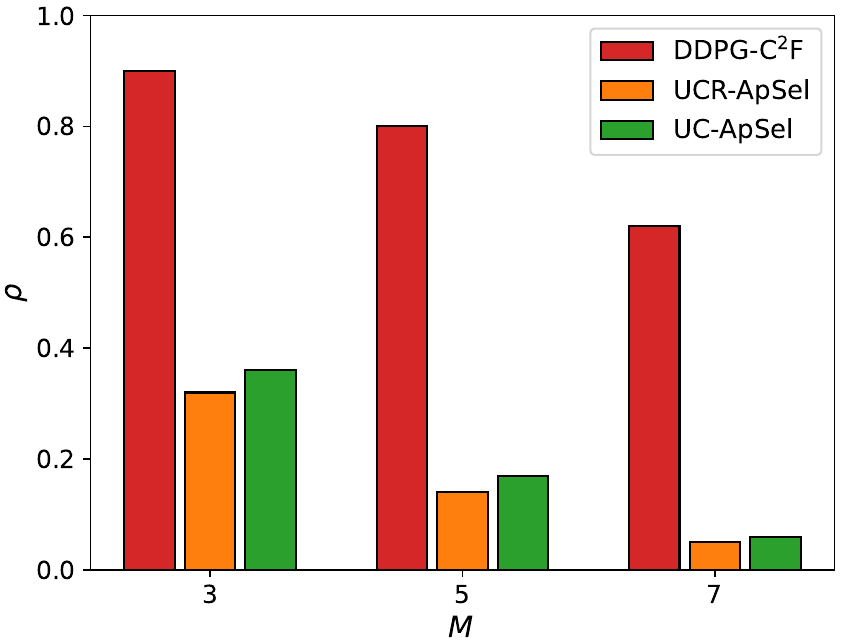}}
\hspace{1mm}
\subfloat[]{
\label{1b}
\includegraphics[width=0.30\textwidth]{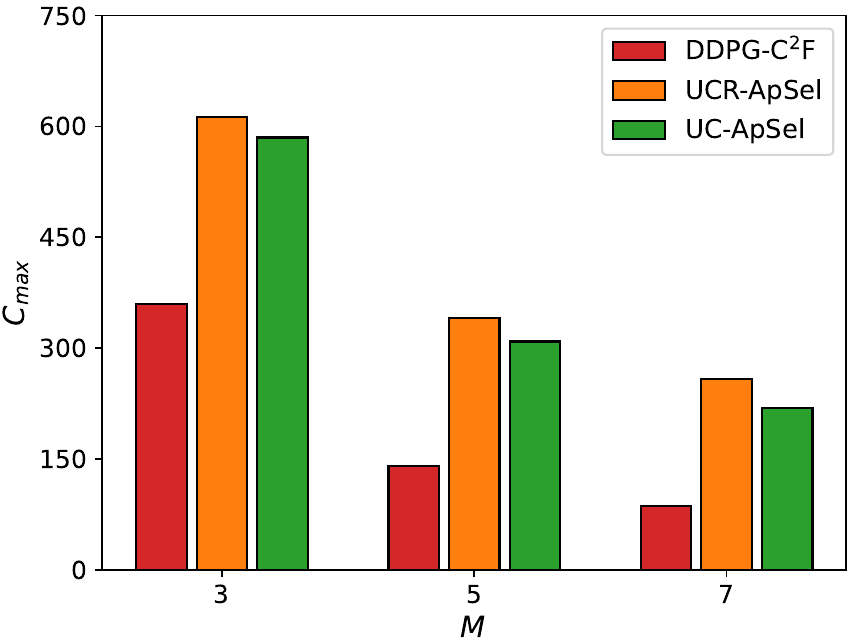}}
\hspace{1mm}
\subfloat[]{
\label{1c}
\includegraphics[width=0.30\textwidth]{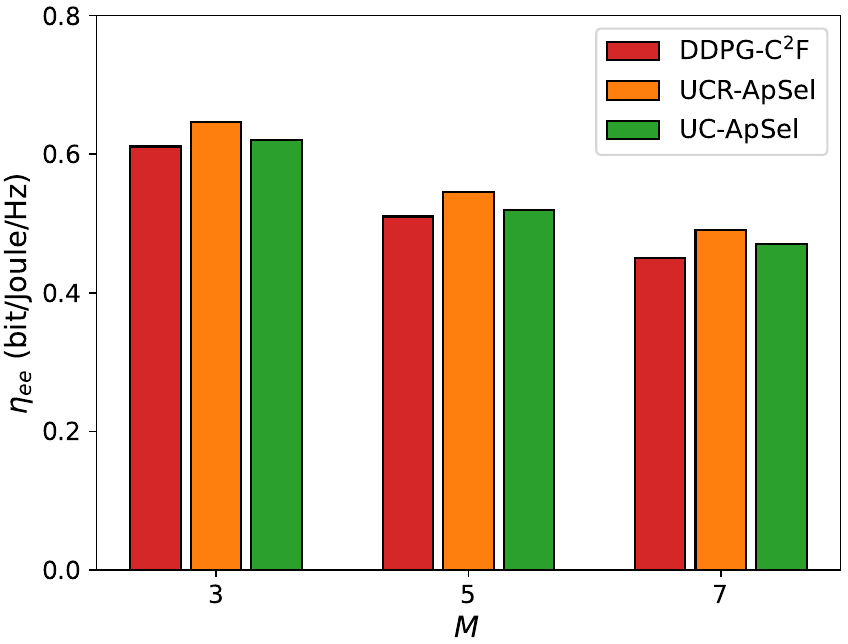}}
\caption{(a) Balance of subnetworks $\rho$, (b) maximum number of channels across all subnetworks $C_{max}$ and (c) energy efficiency $\eta_{ee}$ versus the number of subnetworks $M$ in the case of joint optimization of energy efficiency and balance of subnetworks. $K=50$. $L=100$. $V_{max}=5$m/s.}
\end{figure*}

\subsection{Simulation Results}
To demonstrate the performance of the proposed DDPG-C$^{2}$F framework, we compare our framework with two existing clustered cell-free networking algorithms with AP selection:
\begin{itemize}
\item \textbf{User-centric ratio-fixed AP-selection based clustering (UCR-ApSel)}\cite{10458891}: The UCR-ApSel benchmark selects a given portion of APs after grouping users with hierarchical clustering algorithm, and then assigns each AP to the cluster containing its best user with the highest large-scale fading coefficient. The ratio of the number of APs to the number of users across all subnetworks is fixed at the optimal AP-selection ratio derived in \cite{10458891}.
\item \textbf{User-centric clustering with AP selection (UC-ApSel) }\cite{10458891}: The UC-ApSel benchmark takes the same approach as the UCR-ApSel algorithm and select the same portion of APs yet without fixing the ratio of the number of APs to the number of users in each subnetwork.
\end{itemize}

\subsubsection{Joint Optimization of Energy Efficiency and Balance of Subnetworks} 
Fig. 9 presents the balance-aware energy efficiency $\eta_{ee}$ during the training process with the proposed DDPG-C$^{2}$F framework and the benchmarks. The variant of the proposed framework utilizing all the large-scale fading coefficients is also plotted. Similarly to Fig. 3, we can clearly see from Fig. 9 that the proposed DDPG-C$^{2}$F framework performs the same as its variant, which corroborates its ability to efficiently reduce the channel estimation and computational cost with no performance degradation. Compared to the benchmarks, significantly higher balance-aware energy efficiency is achieved by the proposed DDPG-C$^{2}$F framework.

Fig. 10 compares the balance-aware energy efficiency $\eta_{\rho}$ achieved with the proposed DDPG-C$^{2}$F framework and the benchmarks. It can be seen from Fig. 10 that the proposed framework achieves much higher balance-aware energy efficiency than the baselines in all scenarios. In addition, it can be observed in Fig. 10(a) that the balance-aware energy efficiency of the proposed DDPG-C$^{2}$F framework decreases as the number of users $K$ increases, while the balance-aware energy efficiency with the benchmarks increases. This is because with a large number of users, more APs are selected to guarantee that ZFBF can be applied to cancel out the intra-subnetwork interference, leading to low energy efficiency and thus lower balance-aware energy efficiency with our DDPG-C$^{2}$F framework. For the benchmarks, although the energy efficiency is lower, more balanced subnetworks can be generated with the hierarchical clustering algorithm as the number of users increases, leading to slightly increased balance-aware energy efficiency.

Fig. 11 presents the balance of subnetworks $\rho$, the maximum number of channels across all subnetworks $C_{max}$ and the energy efficiency $\eta_{ee}$ with the proposed DDPG-C$^{2}$F framework and the benchmarks. Similar to the rate-oriented case, the simulation results corroborate the effectiveness of $\rho$ in measuring the balance of subnetworks, and the superiority of the proposed DDPG-C$^{2}$F framework in balancing subnetwork sizes while achieving similar energy efficiency.

\begin{figure}[t]
\centering                             
\includegraphics[width=0.42\textwidth]{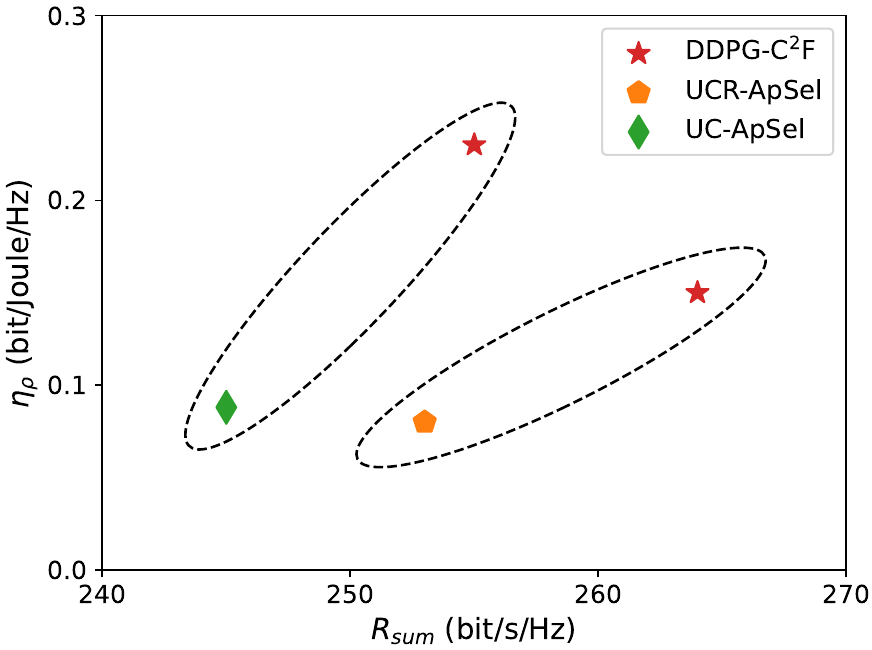}\\
\caption{Balance-aware energy efficiency $\eta_{\rho}$ and sum rate $R_{sum}$ with the proposed DDPG-C$^{2}$F framework and the benchmarks in the case of joint optimization of energy efficiency and balance of subnetworks with sum rate constraint. The results of the proposed framework are obtained by setting the rate threshold $R_{th}$ as the sum rate achieved with the benchmarks in the same dashed circle. $K=50$. $L=100$. $M=5$. $V_{max}=5$m/s.}
\end{figure}

\subsubsection{Joint Optimization of Energy Efficiency and Balance of Subnetworks with Sum Rate Constraint}
Fig. 12 presents the balance-aware energy efficiency $\eta_{\rho}$ and the sum rate $R_{sum}$ in the case with sum rate constraint. Similar to the rate-oriented case, the rate threshold $R_{th}$ is set the same as the sum rate achieved with the benchmark marked in the same dashed circle. It can be seen that the proposed DDPG-C$^{2}$F framework can successfully meet the sum rate requirement, and simultaneously improve the balance-aware energy efficiency compared to the baselines.

\begin{figure}[t]
\centering                             
\includegraphics[width=0.42\textwidth]{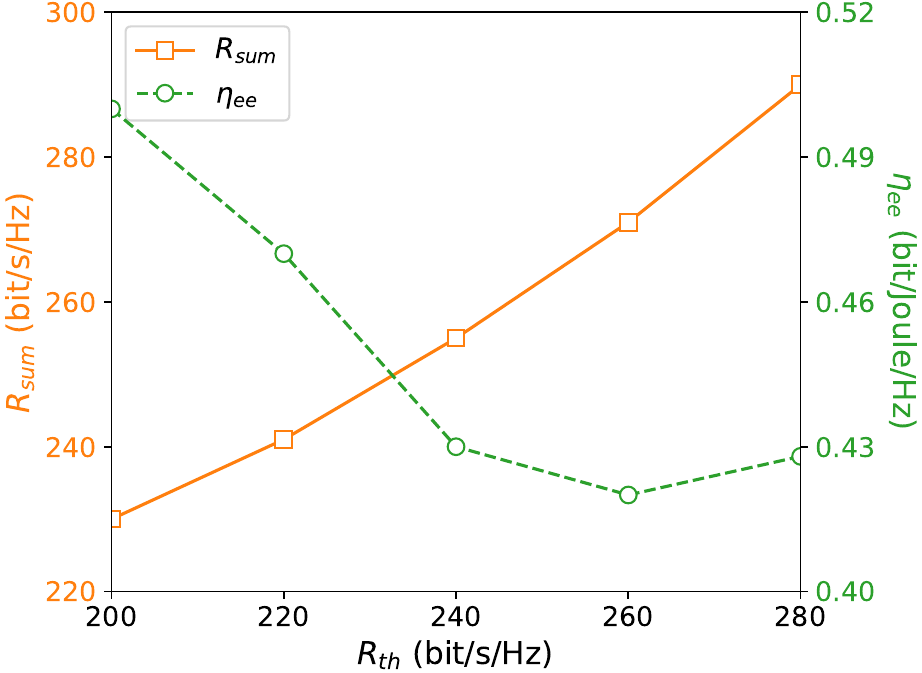}\\
\caption{Sum rate $R_{sum}$ and energy efficiency $\eta_{ee}$ with the proposed DDPG-C$^{2}$F framework under various rate threshold $R_{th}$ in the case of joint optimization of energy efficiency and balance of subnetworks with sum rate constraint. $K=50$. $L=100$. $M=5$. $V_{max}=5$m/s.}
\end{figure}

To have a closer look at the energy efficiency and rate performance, Fig. 13 presents the sum rate $R_{sum}$ and the energy efficiency $\eta_{ee}$ with the proposed DDPG-C$^{2}$F framework by varying rate threshold $R_{th}$. We can see from Fig. 13 that as $R_{th}$ increases, a tradeoff between sum rate and energy efficiency is first observed and then both sum rate and energy efficiency increase. This is because when the rate requirement is relatively low, increasing $R_{th}$ would activate more APs for data transmission, leading to more energy consumption and thus low energy efficiency. After all the APs are activated, keep increasing $R_{th}$ would lead to a giant subnetwork for the sake of meeting the high rate requirement. As sum rate $R_{sum}$ increases with $R_{th}$ while the energy consumed by the APs remains the same, energy efficiency $\eta_{ee}$ also increases.

\subsubsection{Comparison of Computational Complexity and Running Time} 
Table III compares the computational complexity and running time of the proposed DDPG-C$^{2}$F framework with the benchmarks. In the energy efficiency-oriented cases, the dimensionality of the action space is $\left|{\mathcal{A}}\right|=3M$ as given in Section V-A, and hence the computational complexity of the proposed framework is $O(L+3M)=O(L)$. Both the UCR-ApSel and the UC-ApSel benchmarks have the same computational complexity of $O(K^{2}L)$ [16]. Similar to the rate-oriented cases, we can see from Table III that the computational complexity and running time of the proposed DDPG-C$^{2}$F framework is much lower than those of the benchmarks.

\begin{table}[t]
\centering
\caption{Comparison of Computational Complexity and Running Time in Energy Efficiency-Oriented Cases. $M=5$, $K=50$, $L=100$.}
\begin{tabular}{|c|c|c|} \hline
\textbf{Algorithms} & \textbf{Computational complexity} & \textbf{Running time (s)} \\ \hline
DDPG-C$^{2}$F & $O(L)$ & 0.08 \\ \hline
UCR-ApSel & $O(K^{2}L)$ & 0.16 \\ \hline
UC-ApSel & $O(K^{2}L)$ & 0.15 \\ \hline
\end{tabular}
\end{table}

\section{Discussions}
Although the proposed DDPG-C$^{2}$F framework optimizes the long-term performance, the optimal clustered cell-free networking result could vary from one time interval to another due to user mobility. As a result, users need to frequently reassociate with APs, bringing high handover cost. In addition, the number of users is assumed to be fixed in the previous sections, whereas users in practical systems might access or leave the network at any time. In this section, we will show that the proposed DDPG-C$^{2}$F framework is able to reduce the number of handovers compared to the benchmarks and robust in the case with dynamically varying number of users. Moreover, Sections VI-C and VI-D are added to compare our proposed DDPG-C$^{2}$F framework with DRL benchmarks and investigate the sensitivity of our framework to hyperparameters, respectively.

\begin{figure}[t]
\centering
\subfloat[\footnotesize Joint optimization of sum rate and balance of subnetworks]{\includegraphics[width=0.23\textwidth]{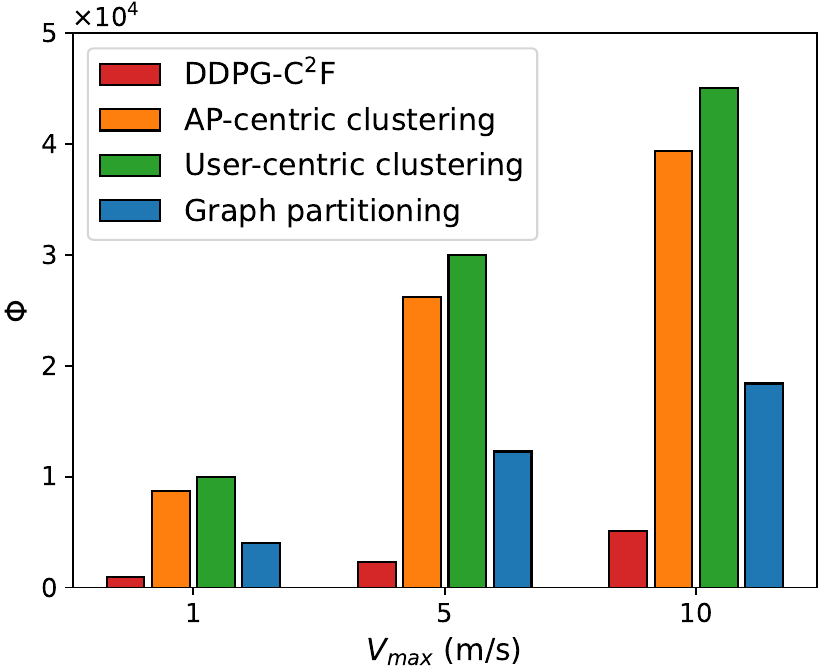}}\hspace{1mm}
\subfloat[\footnotesize Joint optimization of energy efficiency and balance of subnetworks]{\includegraphics[width=0.238\textwidth]{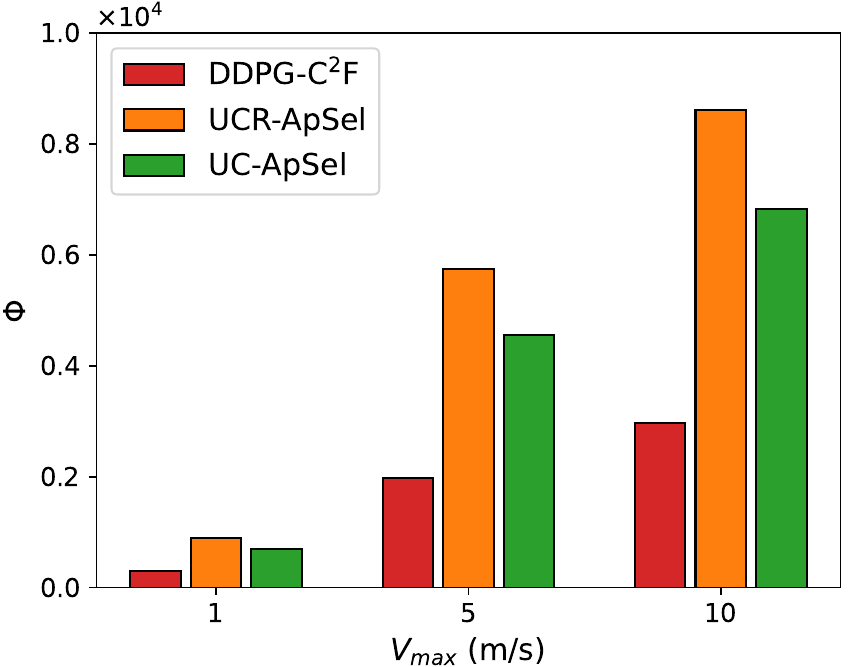}}
\caption{Number of handovers $\Phi$ with the proposed DDPG-C$^{2}$F framework and the benchmarks versus the maximum speed of users $V_{max}$. $K=50$. $L=100$. $M=5$.}
\end{figure}

\subsection{Handover Issues}
Let us define the number of handovers $\Phi^{(t)}$ as the number of new connections established at time interval $t$ compared to those at time interval $t-1$, i.e.,
\begin{equation}
\Phi^{(t)}=\sum _{k=1}^K\sum _{l=1}^L \psi_{k,l}^{(t)}(1-\psi_{k,l}^{(t-1)})\tag{44},
\end{equation}
where $\psi_{k,l}^{(t)}$ is the association indicator. $\psi_{k,l}^{(t)}=1$ if user $u_{k}$ and AP $b_{l}$ are in the same subnetwork, and $\psi_{k,l}^{(t)}=0$, otherwise.

Fig. 14 shows the total number of handovers over the time horizon, $\Phi=\sum _{t=1}^{T}\Phi^{(t)}$, with the proposed DDPG-C$^{2}$F framework and the benchmarks versus the maximum speed of users $V_{max}$. Due to page limit, only the results in the cases without rate constraint are presented for demonstration purpose. It can be seen in Fig.14 that the proposed DDPG-C$^{2}$F framework requires significantly smaller number of handovers than the benchmarks. This indicates that our framework can considerably reduce the transmission delay incurred by handover and improve the service quality over user mobility.
The reason is that the proposed framework partitions a network according to the features of subnetwork anchors learned from the current channel features. As channel features vary smoothly as users move, similar anchor features would be generated in consecutive time intervals. However, the inherent randomness of the benchmarking clustering and graph partitioning algorithms could cause significant variation in network partitioning. Moreover, we can see from the figure that the number of handovers $\Phi$ increases with the maximum speed of users $V_{max}$. It highlights the importance of jointly optimizing handover management and clustered cell-free networking in high-mobility scenarios, which should be carefully studied in the future.

\begin{figure}[t]
\centering
\subfloat[\footnotesize Joint optimization of sum rate and balance of subnetworks]{\includegraphics[width=0.23\textwidth]{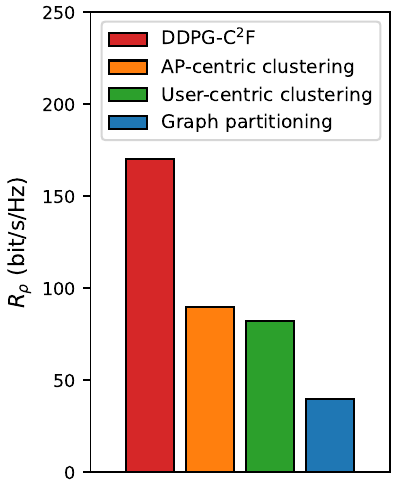}}\hspace{1mm}
\subfloat[\footnotesize Joint optimization of energy efficiency and balance of subnetworks]
{\includegraphics[width=0.23\textwidth]{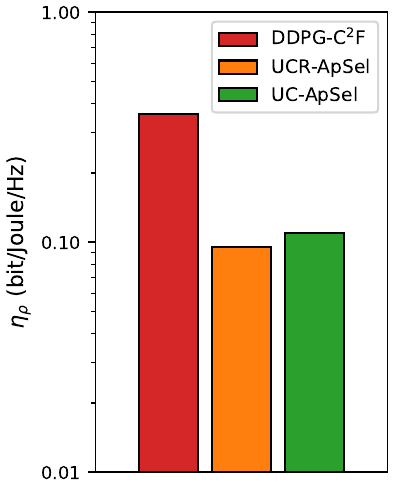}}
\caption{Comparison of the balance-aware sum rate $R_{\rho}$ and the balance-aware energy efficiency $\eta_{\rho}$ in a dynamic scenario with random user accessing and leaving. $L=100$. $M=5$. $V_{max}=5$m/s. $K^{(0)}=50$. $K^{'}_{max}=5$.}
\end{figure}

\subsection{Varying Number of Users}
For the case with users randomly accessing and leaving the network, we model the time-varying number of users as
\begin{equation}
K^{(t)}=K^{(t-1)}+K^{'}\tag{45},
\end{equation}
where $K^{'}\in\left\{-K^{'}_{max}, -K^{'}_{max}+1, \cdots, K^{'}_{max}\right\}$ denotes the change of the number of users in two consecutive time intervals. $K^{'}>0$ indicates that new users access the network, while $K^{'}<0$ represents that existing users leave the network. 

Fig. 15 compares the proposed DDPG-C$^{2}$F framework and the benchmarks in the dynamic scenario with varying number of users. We can observe from Fig. 15 that the proposed framework outperforms the benchmarks, indicating that the proposed DDPG-C$^{2}$F framework can accommodate random user accessing and leaving without the need of learning from scratch. This appealing feature originates from our design that the dimensionality of the state space and the action space is independent of the number of users in the network. Since the number of neurons in the input layer and the output layer of the neural network are determined by the dimensionality of the state space and the action space, respectively, the numbers of neurons in both input and output layers are also independent of the number of users in the network. Therefore, the proposed DDPG-C$^{2}$F framework is robust to the dynamic scenario with a varying number of users.

\subsection{Comparison with DRL Benchmarks}
To further demonstrate the performance of the proposed DDPG-C$^{2}$F framework, we compare it with the following two DRL benchmarks that directly learn the optimal clustered cell-free networking result in this subsection.
\begin{itemize}
\item \textbf{Deep Q-network based clustered cell-free networking (DQN-C$^{2}$F)}: The agent learns the optimal clustered cell-free networking result by evaluating all possible network partitions. The channel features in the state space and the reward function are the same as those in the proposed DDPG-C$^{2}$F framework.
\item \textbf{Multi-agent proximal policy optimization based clustered cell-free networking (MAPPO-C$^{2}$F)}: Each AP acts as an agent to learn which subnetwork it belongs to independently, and then each user is assigned to the subnetwork containing its
closest AP. An AP observes the large-scale fading coefficients between itself and all users, and receives the same reward as the proposed DDPG-C$^{2}$F framework to ensure global coordination.
\end{itemize}
Since the DQN-C$^{2}$F benchmark suffers from the curse of dimensionality as the network scales, we compare the proposed DDPG-C$^{2}$F framework with the DQN-C$^{2}$F benchmark in a small-scale network.

\begin{figure}[t]
\centering
\subfloat[\footnotesize Joint optimization of sum rate and balance of subnetworks]{
\label{1a}
\includegraphics[width=0.23\textwidth]{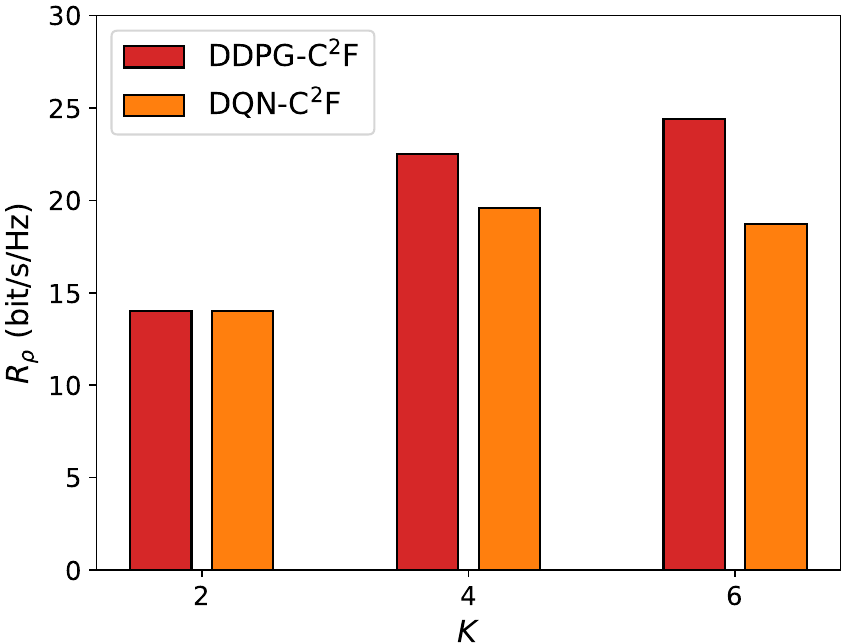}}
\hspace{0.5mm}
\subfloat[\footnotesize Joint optimization of energy efficiency and balance of subnetworks]{
\label{1b}
\includegraphics[width=0.23\textwidth]{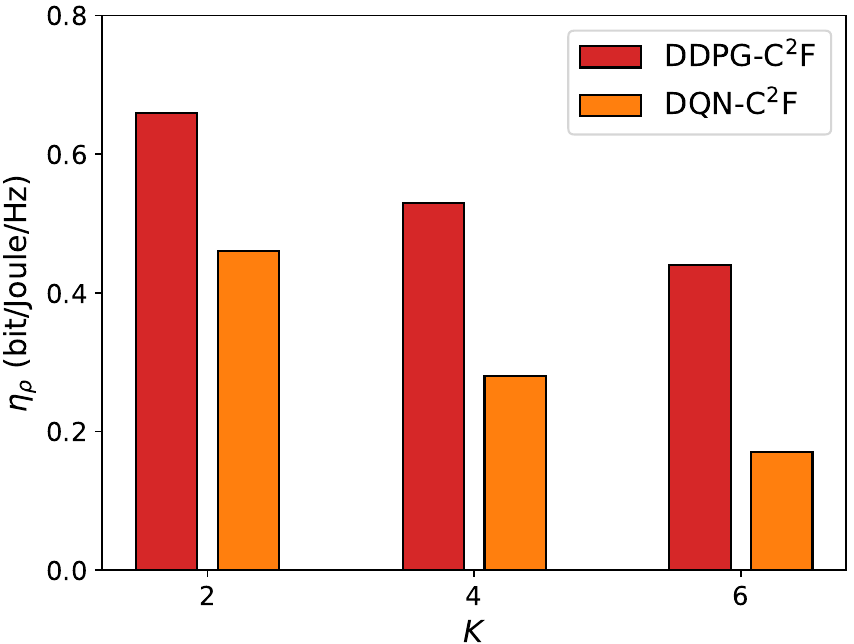}}
\caption{Comparison of the balance-aware sum rate $R_{\rho}$ and the balance-aware energy efficiency $\eta_{\rho}$ with the proposed DDPG-C$^{2}$F framework and the DQN-C$^{2}$F benchmark versus the number of users $K$. $L=8$. $M=2$. $V_{max}=5$m/s.}
\end{figure}

\begin{figure}[t]
\centering
\subfloat[\footnotesize Joint optimization of sum rate and balance of subnetworks]{
\label{1a}
\includegraphics[width=0.23\textwidth]{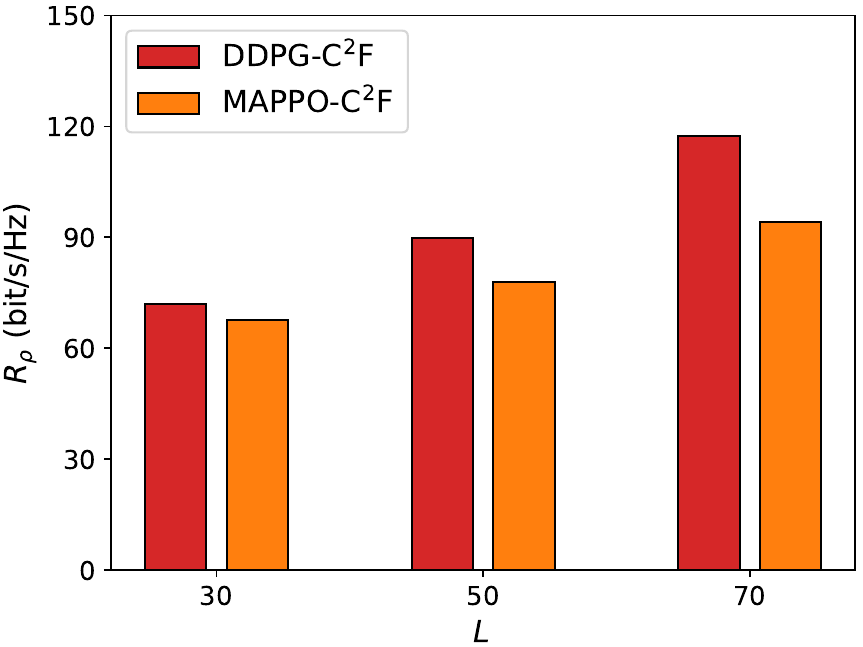}}
\hspace{0.5mm}
\subfloat[\footnotesize Joint optimization of energy efficiency and balance of subnetworks]{
\label{1b}
\includegraphics[width=0.23\textwidth]{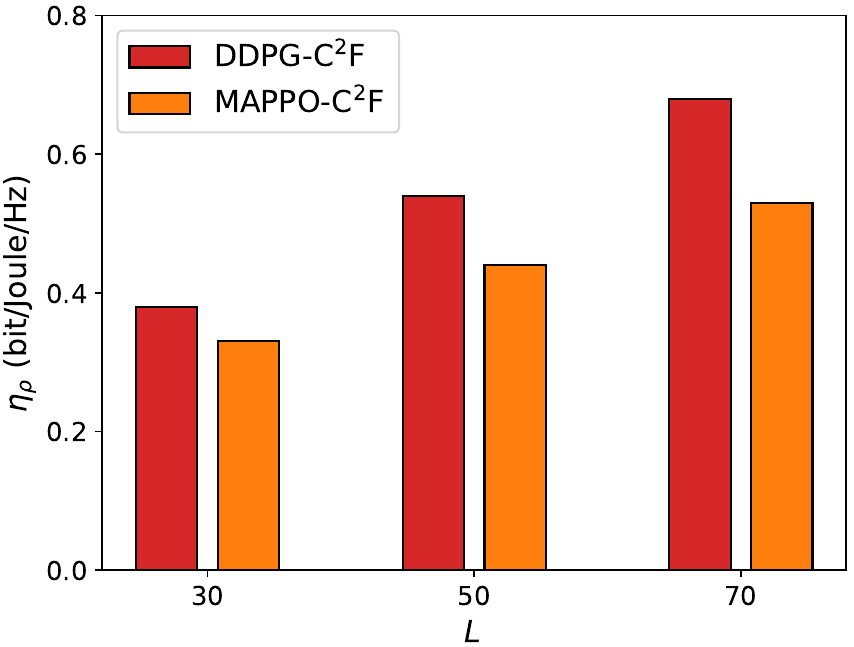}}
\caption{Comparison of the balance-aware sum rate $R_{\rho}$ and the balance-aware energy efficiency $\eta_{\rho}$ with the proposed DDPG-C$^{2}$F framework and the MAPPO-C$^{2}$F benchmark versus the number of APs $L$. $K=20$. $M=3$. $V_{max}=5$m/s.}
\end{figure}

Fig. 16 presents the balance-aware sum rate $R_{\rho}$ and the balance-aware energy efficiency $\eta_{\rho}$ with the proposed DDPG-C$^{2}$F framework and the DQN-C$^{2}$F benchmark. It can be seen from Fig. 16(a) that the proposed DDPG-C$^{2}$F framework outperforms the DQN-C$^{2}$F benchmark when the number of users $K$ is large. This is because the DQN-C$^{2}$F benchmark could fall into local optima with a high-dimensional solution space. Moreover, Fig. 16(b) shows that more performance gains can be achieved in the energy efficiency-oriented case, as incorporating AP selection into clustered cell-free networking would greatly increase the number of possible network partitions, making it difficult for the DQN-C$^{2}$F benchmark to find the optimal solution.

Fig. 17 compares the proposed DDPG-C$^{2}$F framework with the MAPPO-C$^{2}$F benchmark when the number of APs $L=30,~50,$ and $70$. We can see from Fig. 17 that the proposed DDPG-C$^{2}$F framework achieves better performance than the MAPPO-C$^{2}$F benchmark in both rate-oriented and energy efficiency-oriented cases, and the gains become prominent as the number of APs $L$ increases. This is because with MARL, the joint solution space of $L$ agents grows exponentially with $L$, and a large solution space makes it difficult for each agent to learn the optimal policy that maximizes the global objective.

\begin{figure}[t]
\centering
\subfloat[$\sigma_{a, max}=0.25$]{
\label{1a}
\includegraphics[width=0.24\textwidth]{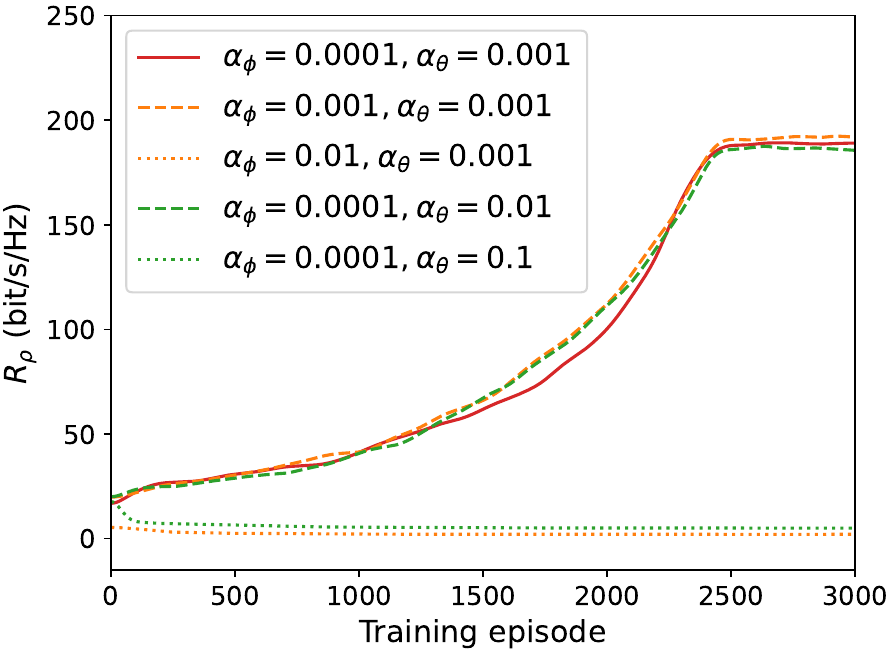}}
\subfloat[$\alpha_{\phi}=0.0001$, $\alpha_{\theta}=0.001$]{
\label{1b}
\includegraphics[width=0.24\textwidth]{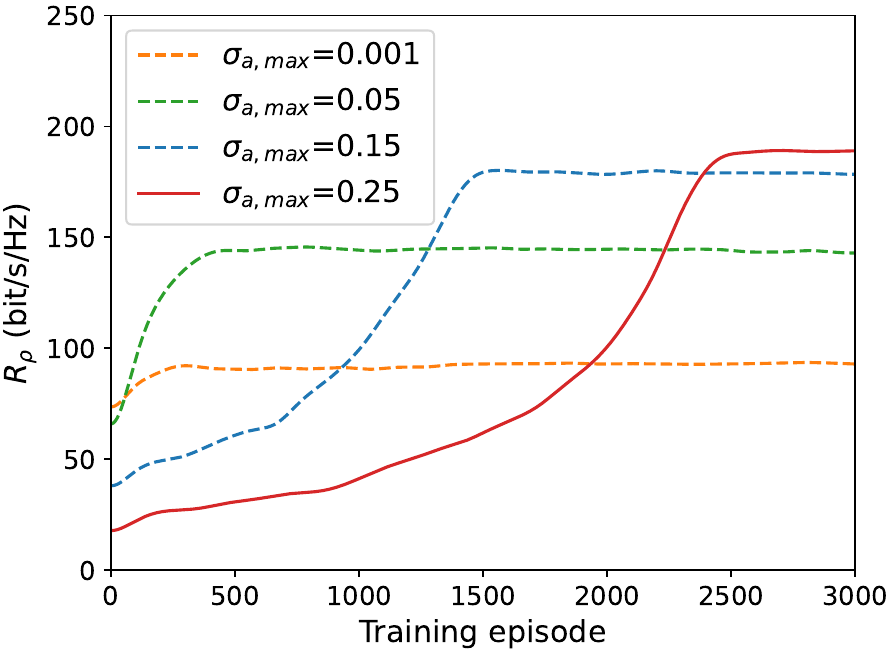}}
\caption{Balance-aware sum rate $R_{\rho}$ during the training process in the case of joint optimization of sum rate and balance of subnetworks with (a) $\sigma_{a, max}=0.25$ and (b) $\alpha_{\phi}=0.0001$, $\alpha_{\theta}=0.001$. $K=50$. $L=100$. $M=5$. $V_{max}=5$m/s.}
\vspace{-1mm}
\end{figure}

\subsection{Training Sensitivity to Hyperparameters}
To investigate the sensitivity of the proposed DDPG-C$^{2}$F framework to key hyperparameters, Fig. 18 presents the training processes of the proposed framework under various values of the learning rate of actor network $\alpha_{\phi}$, the learning rate of critic network $\alpha_{\theta}$, and the maximum standard deviation $\sigma_{a, max}$ of the action noise. Due to page limit, only the results in the case of joint optimization of sum rate and balance of subnetworks are presented. It can be observed from Fig. 18(a) that excessively large learning rates prevent the proposed framework from converging. Moreover, the actor network requires a smaller learning rate than the critic network. This is because a large learning rate for the actor network leads to significant variations in the policy of the agent, thereby degrading the training stability. By contrast, a large learning rate helps the critic network quickly learn the expected cumulative reward of the action taken by the agent. This ensures that the agent updates its policy in the direction of higher reward during the training process. Additionally, it is shown in Fig. 18(b) that by enlarging the maximum standard deviation $\sigma_{a, max}$ of the action noise, the proposed DDPG-C$^{2}$F framework achieves better performance with sufficient training, as the agent can explore the solution space more thoroughly with stronger action noise.

\section{Conclusion}
This paper studied the clustered cell-free networking problem over user mobility. To promptly produce the optimal network partition as users move, we delved into the DRL approach and proposed a DDPG-C$^{2}$F framework. In contrast to the existing works that require the channel conditions between all users and all APs, our framework only requires one single channel estimate at each AP, which reduces not only the channel estimation costs, but also the training and inference costs due to the reduced dimensionality of the state space. Moreover, the proposed framework can be adopted in various application scenarios to solve different clustered cell-free networking problems with different objectives and constraints. In addition, our framework can reduce the handover cost incurred by user mobility, and is able to accommodate the dynamic scenario with random user accessing or leaving. Simulation results demonstrated the effectiveness of the proposed framework in four cases along with its superior performance over the benchmarks. The significant performance gains achieved with our framework in all scenarios showcase the potential of the proposed DRL-based clustered cell-free networking framework for practical implementation.

\bibliographystyle{IEEEtran}
\bibliography{reference}

\begin{IEEEbiography}[{\includegraphics[width=1in,height=1.25in,clip,keepaspectratio]{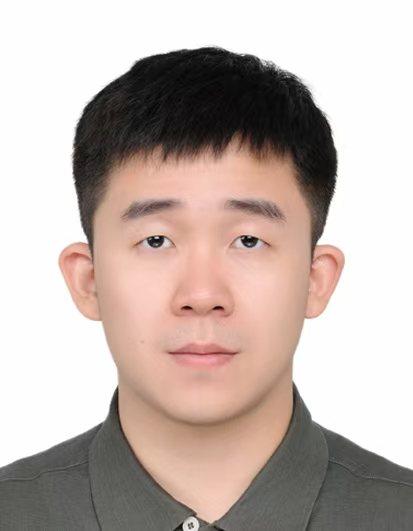}}]{Ouyang Zhou}\enspace received the B.S. degree from Nankai University, Tianjin, China, in 2019. He is currently pursuing the Ph.D. degree with the College of Electronic and Information Engineering, Tongji University, Shanghai China. His current research interests include clustered cell-free networking and deep reinforcement learning. 
\end{IEEEbiography}

\begin{IEEEbiography}[{\includegraphics[width=1in,height=1.25in,clip,keepaspectratio]{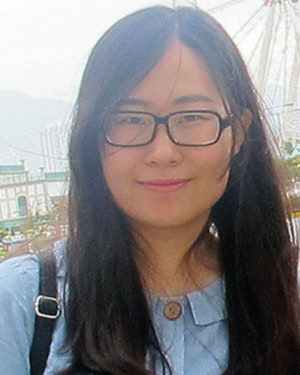}}]{Junyuan Wang}\enspace (Member, IEEE) received the B.S. degree in Communications Engineering from Xidian University, Xi'an, China, in 2010, and the Ph.D. degree in Electronic Engineering from City University of Hong Kong, Hong Kong, China, in 2015. From 2015 to 2017, she was a Research Associate in the School of Engineering and Digital Arts, University of Kent, Canterbury, U.K. From 2018 to 2020, she was a Lecturer (Assistant Professor) in the Department of Computer Science, Edge Hill University, Ormskirk, U.K. She is currently a Research Professor with the College of Electronic and Information Engineering and the Institute for Advanced Study, Tongji University, Shanghai, China. Her research mainly focuses on wireless communications and networking, and  artificial intelligence. She was a co-recipient of the Best Paper Award from IEEE International Conference on Communications in China (ICCC) in 2024 and the Best Student Paper Award from IEEE 85th Vehicular Technology Conference–Spring (VTC-Spring) in 2017.
\end{IEEEbiography}

\begin{IEEEbiography}[{\includegraphics[width=1in,height=1.25in,clip,keepaspectratio]{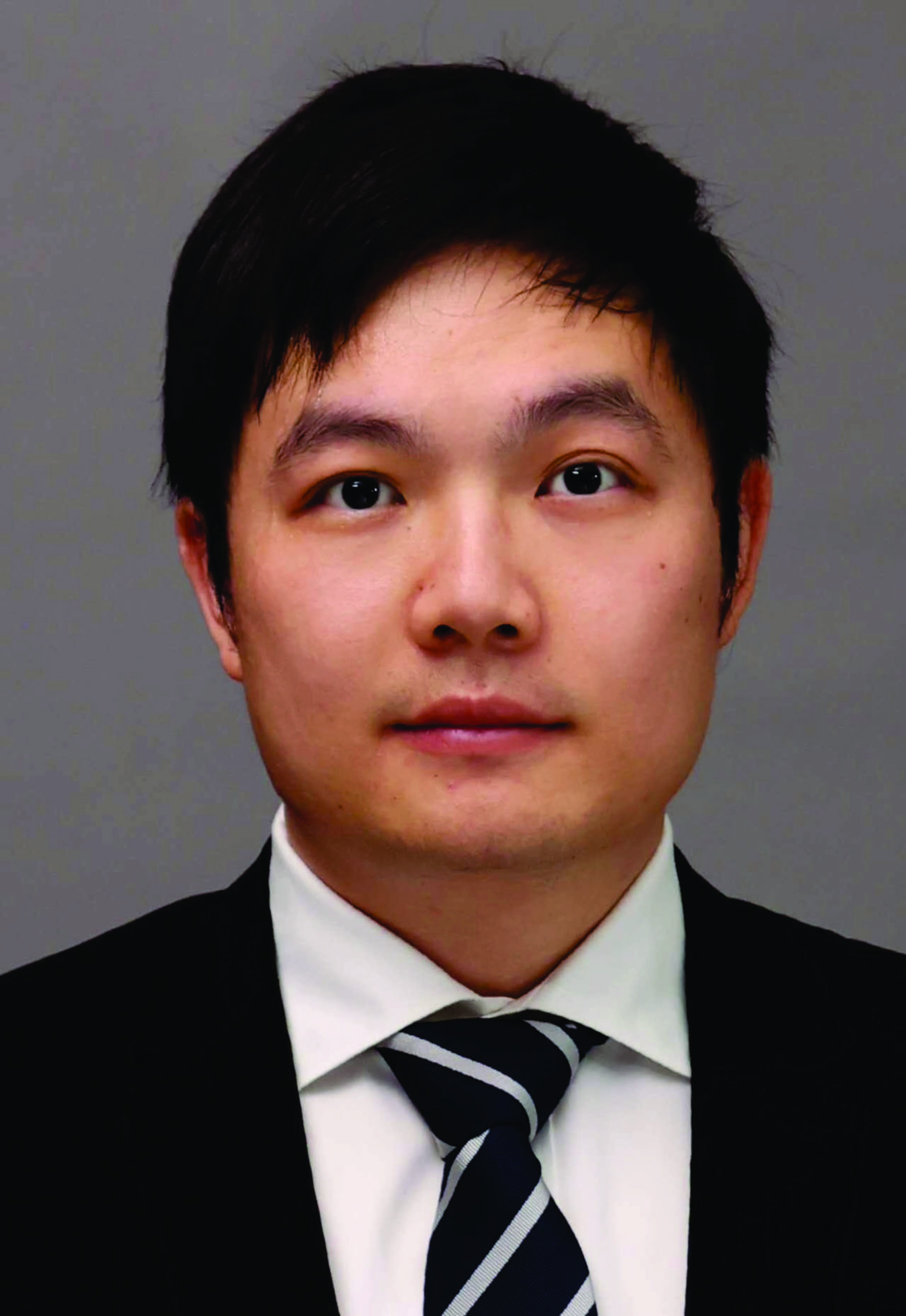}}]{Bo Qian}\enspace (Member, IEEE) received the B.S. and M.S. degrees from the College of Mathematics at Sichuan University, Chengdu, China, in 2015 and 2018, respectively, and the Ph.D. degree in Information and Communication Engineering at Nanjing University, Nanjing, China, in 2022. From 2022 to 2024, he was a Postdoctoral Fellow with the Peng Cheng Laboratory, Shenzhen, China. From 2024 to 2026, he worked as a Researcher and Assistant Professor (Special Appointment) at the Information Systems Architecture Science Research Division, National Institute of Informatics, Tokyo, Japan. From 2026, he is an Assistant Professor (Special Appointment) with the Graduate School of Information Science and Technology, The University of Tokyo, Japan. His research interests include cellular RAN, SAGIN, vehicular networks, and the industrial Internet. He received the Best Paper Award at IEEE VTC2020-Fall.
\end{IEEEbiography}

\begin{IEEEbiography}[{\includegraphics[width=1in,height=1.25in,clip,keepaspectratio]{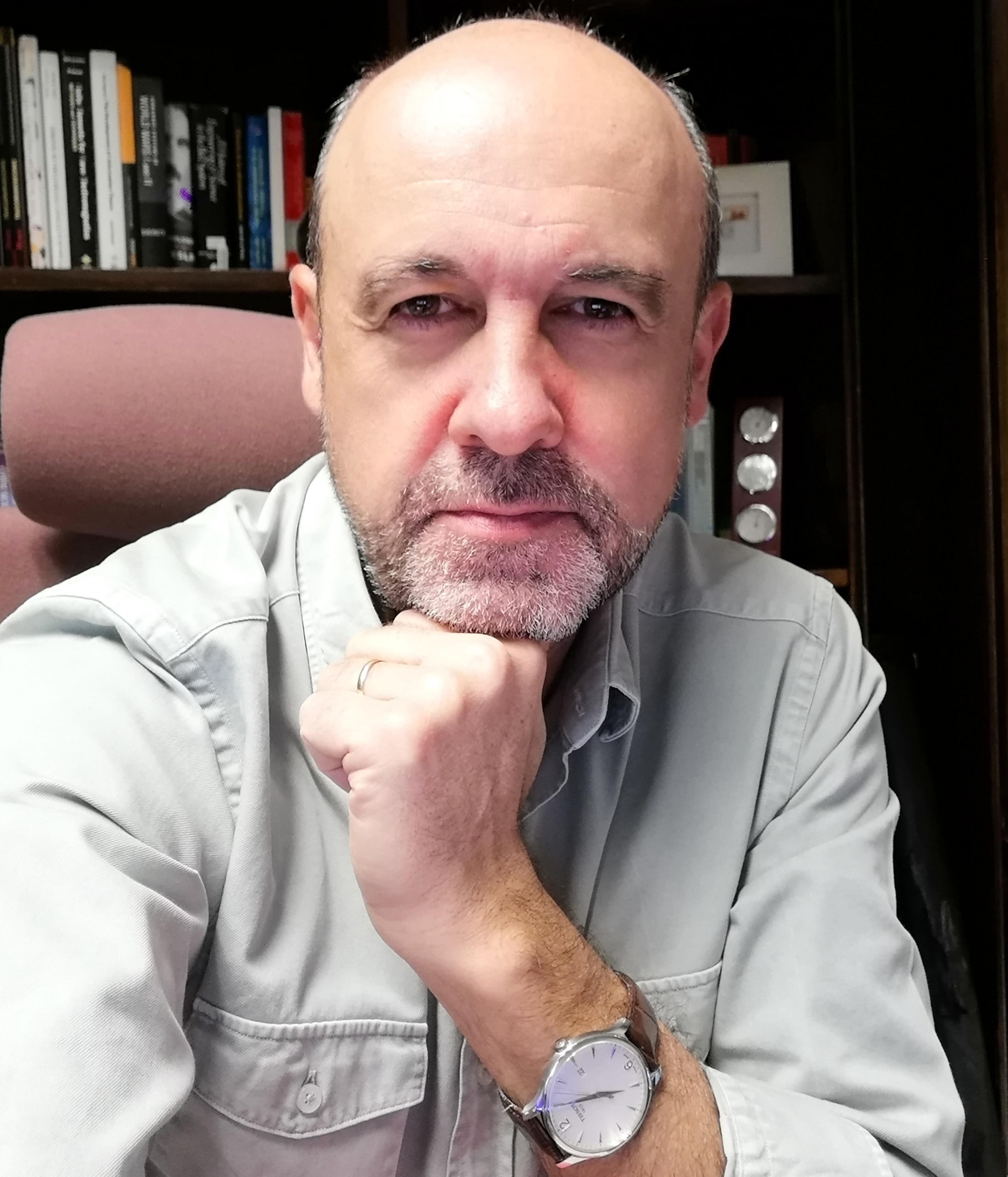}}]{Antonio~P\'{e}rez~Yuste}\enspace (Senior Member, IEEE) received the B.Eng. degree in radio communications, the M.Eng. degree in telecommunications, and the Ph.D. degree (cum laude) in telecommunications engineering, all from the Universidad Politécnica de Madrid (UPM), Spain, in 1991, 1996, and 2004, respectively. He is currently a Tenured Professor in the Department of Communications and Audio and Video Engineering at UPM, where his research interests focus on long-range and short-range wireless communication systems. Prof. Pérez Yuste has held several leadership roles at UPM: Vice Director of the School of Telecommunication Systems Engineering (1997–2001), Director of the same school (2001–2004), Assistant to the Rector (2004–2012), Director of the UPM Sino-Spanish Campus in Shanghai, China (September 2012 – February 2014), and Head of the UPM Sino-Spanish Cooperation Office in Madrid (March 2014 – July 2014). In 2014, he was appointed to a three-year Guest Professorship at Tongji University in Shanghai under the Chinese Government’s prestigious National High-End Foreign Expert Program. From 2017 to 2019, he continued as a Guest Professor at the same university’s College of Electronics and Information Engineering in Shanghai. He received the UPM Medal in 2004 for his outstanding work as School Director, the UPM Prize for Innovation in Education in 2009 for his best educational practices, and the UPM Prize for Excellence in Education in 2018 for his outstanding academic career. In December 2020, he was awarded the ‘Remón y Zarco del Valle’ Prize given in the UPM to the best international research paper on security and defense.
\end{IEEEbiography}

\begin{IEEEbiography}[{\includegraphics[width=1in,height=1.25in,clip,keepaspectratio]{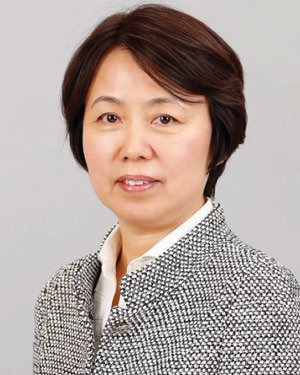}}]{Yusheng Ji}\enspace (Fellow, IEEE) received the B.E., M.E., and Ph.D. degrees in electrical engineering from The University of Tokyo. She joined the National Center for Science Information Systems (NACSIS), Tokyo, Japan, in 1990, as an Assistant Professor and then became an Associate Professor. In 2000, she was an Associate Professor and then a Professor with the National Institute of Informatics (NII), Tokyo, and has been with the Graduate University for Advanced Studies, SOKENDAI, Japan, since 2002. She has been a Professor Emeritus with NII and SOKENDAI since April 2026. She has published more than 600 refereed papers in the areas of network resource management, traffic control and performance evaluation, and mobile computing. She is a Distinguished Speaker of the IEEE Vehicular Technology Society. She has received numerous paper awards, including the IEEE Andrew P. Sage Best Transactions Paper Award and the IEEE Communications Society Outstanding Paper Award. She is the General Chair of IEEE INFOCOM 2026. She has served as the TPC Co-Chair, the Symposium Co-Chair, and the Track Co-Chair for major conferences, such as IEEE INFOCOM, ICC, GLOBECOM, and VTC. She is serving/has served editorship for transactions and magazine of IEEE, as well as IEICE and IPSJ.
\end{IEEEbiography}

\end{document}